\documentclass[11pt,a4paper]{article}
\usepackage{jheppub} 
\usepackage{nameref}
\usepackage{dsfont}
\usepackage{mathrsfs, amssymb, amsmath, amsfonts, txfonts, latexsym, graphicx}  
\usepackage[utf8]{inputenc}
\usepackage{soul}
\usepackage{physics}
\usepackage{accents}
\usepackage[T1]{fontenc}

\newcommand{\tgamma}{\tilde{\gamma}}
\newcommand{\tzeta}{\tilde{\zeta}}

\newcommand{\intpE}[2]{\int \dd\Omega_{#1,#2}\,}

\makeatletter
\DeclareRobustCommand{\loplus}{\mathbin{\mathpalette\dog@lsemi{+}}}

\usepackage{float}
\usepackage{mathtools}
\newcommand{\defeq}{\vcentcolon=}

\newcommand{\dog@lsemi}[2]{\dog@semi{#1}{#2}{270,90}}
\newcommand{\dog@semi}[3]{%
  \begingroup
  \sbox\z@{$\m@th#1#2$}%
  \setlength{\unitlength}{\dimexpr\ht\z@+\dp\z@\relax}%
  \makebox[\wd\z@]{\raisebox{-\dp\z@}{%
    \begin{picture}(1,1)
    \linethickness{\variable@rule{#1}}
    \roundcap
    \put(0.5,0.5){\makebox(0,0){\raisebox{\dp\z@}{$\m@th#1#2$}}}
    \put(0.5,0.5){\arc[#3]{0.5}}
    \end{picture}%
  }}%
  \endgroup
}
\newcommand{\variable@rule}[1]{%
  \fontdimen8  
  \ifx#1\displaystyle\textfont3\else
    \ifx#1\textstyle\textfont3\else
      \ifx#1\scriptstyle\scriptfont3\else
        \scriptscriptfont3\relax
  \fi\fi\fi
}
\makeatother
\usepackage{tikz-cd}

\usetikzlibrary{arrows.meta,calc}

\usepackage{cleveref}
\usepackage{pict2e}
\usepackage{diagbox}

\newcommand{\beq}{\begin{eqnarray}}
\newcommand{\eeq}{\end{eqnarray}}
\newcommand{\beqn}{\begin{eqnarray}}
\newcommand{\eeqn}{\end{eqnarray}}

\newcommand{\RR}{\mathbb{R}}

\newcommand{\spl}[1]{\mathrm{SL}(#1,\mathbb{R})}

\usepackage{pict2e}

\newcommand{\updown}[2]{^{#1}_{\phantom{#1}#2}}
\newcommand{\downup}[2]{_{#1}^{\phantom{#1}#2}}

\newcommand{\chkM}{{\color{red} \,\checkmark\kern-5pt{}_{M}}}

\newcommand{\ee}{\end{equation}}
\newcommand{\bea}{\begin{eqnarray}}
\newcommand{\eea}{\end{eqnarray}}

\usepackage{BOONDOX-cal}

\usepackage{scalerel}
\usepackage{stackengine,wasysym}

\newcommand\reallywidetilde[1]{\ThisStyle{%
  \setbox0=\hbox{$\SavedStyle#1$}%
  \stackengine{-.1\LMpt}{$\SavedStyle#1$}{%
    \stretchto{\scaleto{\SavedStyle\mkern.2mu\AC}{.5150\wd0}}{.6\ht0}%
  }{O}{c}{F}{T}{S}%
}}

\newcommand{\ldv}[1]{\mathcal{L}_{#1}}

\newcommand{\pairing}[2]{\langle #1\, , \, #2 \rangle}

\newenvironment{Align}{\begin{equation}
\begin{aligned}}
{\end{aligned}
\end{equation}}
\newenvironment{Align*}{\begin{equation*}
\begin{aligned}}
{\end{aligned}
\end{equation*}\par}

\usepackage{color}

\usepackage{layout}
\usepackage{hyperref} 

\DeclareFontFamily{OT1}{rsfs}{}
\DeclareFontShape{OT1}{rsfs}{m}{n}{ <-7> rsfs5 <7-10> rsfs7 <10->rsfs10}{} 
\DeclareMathAlphabet{\mycal}{OT1}{rsfs}{m}{n}

\DeclareFontFamily{OMS}{cmsy}{}
\DeclareFontShape{OMS}{cmsy}{m}{n}{ <5><6><7><8><9> gen * cmsy <10><10.95><12><14.4><17.28><20.74><24.88> cmsy10 }{}
\DeclareMathAlphabet{\mycaltwo}{OMS}{cmsy}{m}{n}
\DeclareFontFamily{U}{MnSymbolC}{}
\DeclareSymbolFont{MnSyC}{U}{MnSymbolC}{m}{n}
\DeclareFontShape{U}{MnSymbolC}{m}{n}{
    <-6>  MnSymbolC5
   <6-7>  MnSymbolC6
   <7-8>  MnSymbolC7
   <8-9>  MnSymbolC8
   <9-10> MnSymbolC9
  <10-12> MnSymbolC10
  <12->   MnSymbolC12}{}
\DeclareMathSymbol{\intprod}{\mathbin}{MnSyC}{'270}
\title{Entanglement Entropy of Quantum Corners}
\author[a]{Luca Ciambelli}
\author[b,c]{, Jerzy Kowalski-Glikman}
\author[b]{, Ludovic Varrin}
\affiliation[a]{Perimeter Institute for Theoretical Physics, 31 Caroline St. N., Waterloo ON, Canada, N2L 2Y5}
\affiliation[b]{National Centre for Nuclear Research, Pasteura 7, 02-093 Warsaw, Poland}
\affiliation[c]{Faculty of Physics and Astronomy, University of Wroclaw, Pl. Maksa Borna 9, 50-204
Wroclaw}
\emailAdd{ciambelli.luca@gmail.com}
\emailAdd{jerzy.kowalski-glikman@uwr.pl.edu}
\emailAdd{ludovic.varrin@ncbj.gov.pl}

\begin{document}
\abstract{In gravitational theories with boundaries, diffeomorphisms can become physical and acquire a non-vanishing Noether charge. Using the covariant phase space formalism, on shell of  the gravitational constraints, the latter localizes on codimension-$2$ surfaces, the corners. The corner proposal asserts that these charges, and their algebras, must be  important ingredients of any quantum gravity theory. In this manuscript, we continue the study of quantum corner symmetries and algebras by computing the entanglement entropy and quantum informational properties of quantum states abiding to the quantum representations of corners in the framework of $2$-dimensional gravity. We do so for two classes of states: the vacuum and coherent states, properly defined. We then apply our results to JT gravity, seen as the dimensional reduction of $4$d near extremal black holes. There, we demonstrate that the entanglement entropy of some coherent quantum gravity states -- states admitting a semiclassical description -- scales like the dilaton, reproducing the semiclassical area law behavior and further solidifying the quantum informational nature of entropy of quantum corners. We then study general states and their gluing procedure, finding a formula for the entanglement entropy based entirely on the representation theory of $2$d quantum corners.}
\maketitle

\newpage

\section{Introduction}

Over the past few decades, the quantum gravity community has drawn inspiration from two major themes. The first is the holographic principle, proposed by Susskind and ’t~Hooft \cite{tHooft:1993dmi,susskind1994}, which hypothesizes that gravitational physics in the bulk of a spacetime can be reconstructed from a "hologram" residing on its asymptotic boundary. This idea found concrete realization in the AdS/CFT correspondence \cite{maldacena1999,witten1998,gubser1998}, which posits that string theory (and thus quantum gravity) on the background $\text{AdS}^5\times S^5$ is equivalent to $\mathcal{N}=4$ super-Yang--Mills theory defined on the asymptotic boundary of $\text{AdS}^5$.

This proposed equivalence between bulk quantum gravity in asymptotically AdS spacetimes and conformal field theories (CFTs) on their asymptotic boundaries has since been extensively studied in various dimensions. Over time, increasing evidence has suggested that the  correspondence is not limited to the original AdS/CFT setting, but instead applies more broadly and may capture a deep and general feature of quantum gravity.

This naturally raises the question: could a similar correspondence also hold for finite regions of spacetime and their boundaries—not just at asymptotic infinity? A concrete realization of this idea is the so-called corner proposal \cite{Donnelly:2016auv,Speranza:2017gxd,Geiller:2017whh,Gomes:2018dxs,Geiller:2019bti,Freidel:2019ees,Carrozza:2022xut,Donnelly:2020xgu,Freidel:2020xyx,Freidel:2020svx,Freidel:2020ayo,Donnelly:2022kfs,Speranza:2022lxr,Joung:2023doq,Ciambelli:2021vnn,Ciambelli:2022vot, Ciambelli:2023bmn, Freidel:2023bnj}.\footnote{In parallel to our approach, a recent complementary line of research \cite{Bousso:2022hlz, Bousso:2023sya, Bousso:2024ysg, Bousso:2024iry, Bousso:2025mdp, Bousso:2025joj, Bousso:2025fgg} has explored the construction of holographic duals for local subregions, uncovering rich quantum informational features.} The key insight here is that, although general covariance precludes the existence of local observables in gravity, one can define Noether charges associated with co-dimension two surfaces—known as corners and denoted by $S$ in this paper—which form the boundaries of finite regions. This can be understood as a direct consequence of Noether's second theorem.

It turns out that the maximal Poisson bracket algebra of these charges, known as the Extended Corner Symmetry Algebra, is both well-defined and universal \cite{Donnelly:2016auv, Speranza:2017gxd, Ciambelli:2021vnn, Freidel:2021cjp}. Its associated symmetry group is a semidirect product of the diffeomorphism group of the corner surface $S$ and, at each point on the corner, a copy of the semidirect product $\mathrm{SL}(2, \mathbb{R}) \ltimes \mathbb{R}^2$.
\begin{align}
      \mathrm{ECS} = \mathrm{Diff}(S) \ltimes \qty(\mathrm{SL}\qty(2,\RR)\ltimes \RR^2)^S
\end{align}
The Extended Corner Symmetry (ECS) group is universal and independent of the specific region that the corner bounds. This is not surprising, because, as a result of diffeomorphism symmetry any two regions (of the same topology) are physically equivalent. It is therefore natural to propose that the ECS group encodes fundamental information relevant to both classical and quantum gravity. In this sense, the ECS group assumes a role analogous to that of the Poincaré symmetry group in (quantum) field theory on flat spacetimes.

By analogy with the Poincaré group, whose representation theory underpins much of particle physics, it is reasonable to expect that the representation theory of the ECS group will play a central role in the structure of quantum gravity. It is sometimes said that particle physics is, at its core, the study of the representation theory of the Poincaré group. Following this perspective, we propose as a foundational hypothesis that aspects of quantum gravity can be understood through the representation theory of the Extended Corner Symmetry group.

In recent papers \cite{Donnelly:2020xgu, Donnelly:2022kfs, Ciambelli:2022cfr, Ciambelli:2024qgi, Varrin:2024sxe, Neri:2025fsh} the foundations of this research program were laid down. In particular, in \cite{Ciambelli:2024qgi, Varrin:2024sxe, Neri:2025fsh}, the projective representation theory of the ECS in the case of two-dimensional gravity was analyzed. In this case the corner becomes just one point and the Extended Corner Symmetry centrally extends to the quantum corner symmetry group $\mathrm{QCS} = \reallywidetilde{\mathrm{SL}}\qty(2,\RR)\ltimes H_3$. A brief review of the derivation of this group and its representations, following the lines of \cite{Ciambelli:2024qgi}, is provided in section \ref{sec:quantumcorners}.

The second major theme that shaped the development of theoretical physics in the last half of the century is the continuous effort to understand the origin of the Bekenstein-Hawking entropy of black holes, especially the unique feature of this entropy being proportional to the black hole area. It was realized in \cite{Sorkin:1984kjy,Bombelli:1986rw,Srednicki:1993im,Jacobson:1994iw} that the Bekenstein-Hawking thermodynamical entropy is actually a more fundamental entropy of entanglement, that is, the von Neumann entropy associated with the reduced density matrix obtained by tracing out the degrees of freedom inside of the black hole.

Entanglement entropy, which quantifies the relationship between a system and its surroundings (i.e., complement), plays a natural and important role in understanding the microscopic structure of physical systems. A natural first step to investigate this relationship is to decompose the system into subsystems and investigate how to consistently cut and glue different regions. For $2$d gravity, this aspect of the problem was addressed in our previous work \cite{Ciambelli:2024qgi}.

In algebraic quantum field theory, one associates an algebra of operators to a region of spacetime. Thus, decomposing a system into subsystems corresponds to studying the local algebras associated with these subregions (see \cite{Haag:1996hvx,Harlow:2016vwg,Witten:2018zxz,Geng:2021hlu,Chandrasekaran:2022cip,Chandrasekaran:2022eqq,Klinger:2023auu,AliAhmad:2024eun,Sorce:2023fdx, Klinger:2023tgi,Jensen:2023yxy,Faulkner:2024gst, Geng:2025rov,Caminiti:2025hjq,Geng:2025bcb}, for in depth discussion). However, a fundamental obstacle arises in continuum quantum field theories: the algebra of operators does not factorize into subalgebras corresponding to local regions. This issue stems from the universal nature of ultraviolet divergences in quantum field theory, which prevents one from partitioning spacetime without generating infinite entanglement at the entangling surface. Technically, the algebra associated with a subregion is a von Neumann algebra of type $III_1$.

To circumvent this complication, let us assume that we are dealing with a finite-dimensional quantum mechanical system, or equivalently, that the quantum field theory has been placed on a lattice. In such cases, given two subregions $A$ and $B$, the Hilbert space factorizes as
\beq\label{split}
H = H_A \otimes H_B.
\eeq
For a state $\ket{\psi} \in H$, the reduced density matrix is obtained tracing out the degrees of freedom in $B$:
\beq
\rho_A = \mathrm{Tr}_B \ket{\psi} \bra{\psi},
\eeq
leading to the von Neumann entanglement entropy associated with region $A$:
\beq\label{svn}
S_A = -\mathrm{Tr}(\rho_A \log \rho_A).
\eeq
This quantity measures the entanglement between subregion $A$ and its complement $A^c = B$, or equivalently, how mixed the state is within $A$ after tracing over $B$. In a continuum QFT, however, the entanglement entropy \eqref{svn} becomes formally divergent, reflecting the breakdown of the factorization property \eqref{split} and indicating infinite quantum correlations across subregions. This divergence highlights the utility of entanglement entropy as a diagnostic tool for understanding the quantum structure of subregions.

The situation becomes even more subtle in gauge theories, where gauge constraints are inherently non-local, entangling different subregions and complicating any attempt at algebraic factorization. However, studying the behavior of symmetries in the presence of boundaries suggests a potential resolution. In particular, as a consequence of Noether’s second theorem, gauge symmetries can give rise to codimension-2 localized charges at the edges of a boundary—commonly referred to as corner charges. These charges can be canonically incorporated into the phase space through a consistent extension that includes additional edge degrees of freedom. The resulting extended phase space becomes a suitable subspace of the tensor product of the two spaces, obtained by quotienting out the gauge directions. This enables a consistent cut-and-glue procedure in gauge theories that respects the gauge constraints across interfaces.

Entanglement entropy can thus be meaningfully associated with an arbitrary bounded region \cite{Casini:2011kv,Wong:2017pdm,Lin:2018bud,David:2022jfd,Mukherjee:2023ihb,Caminiti:2024ctd,Soni:2024aop, Fliss:2025kzi}. This entails the appearance of edge modes at the corner, which contribute to the entanglement entropy \cite{Ball:2024hqe,Ball:2024xhf,Ball:2024gti}, a feature that can be  directly detected from the gravitational path integral \cite{Anninos:2020hfj, Grewal:2022hlo, Kapec:2024zdj, Law:2025ktz, Law:2025yec}. These edge modes are at the nexus of the connection between gravitational phase spaces (Hilbert spaces) and (quantum) reference frames, as recently explored in \cite{AliAhmad:2021adn, Carrozza:2022xut, Goeller:2022rsx, Klinger:2023tgi, AliAhmad:2023etg, Araujo-Regado:2024dpr, AliAhmad:2024wja, AliAhmad:2024eun, AliAhmad:2024vdw, Kirklin:2024gyl, DeVuyst:2024khu, Geng:2024dbl, Araujo-Regado:2025ejs}. As discussed, a universal algebraic structure—the Extended Corner Symmetry (ECS)—is naturally associated with the region’s boundary. It is therefore compelling to explore how much information about entanglement entropy can be extracted purely from the algebraic structure of the ECS group. This investigation forms the central focus of the present paper. 

Following our previous work \cite{Ciambelli:2024qgi}, we continue here to explore the two-dimensional case. While such models may initially appear to be of purely academic interest, two-dimensional gravity plays a physically meaningful role as a dimensional reduction of four-dimensional gravity in spherically symmetric settings. Many spacetimes of interest—such as Rindler space, causal diamonds, Schwarzschild black holes, and (Anti-)de Sitter space—fall within this class and can thus be effectively studied using the methods developed here.

In this paper, we specifically employ Jackiw–Teitelboim (JT) gravity \cite{Jackiw:1984je,teitelboim1983gravitation} as our working model (see \cite{Mertens:2022irh} for a review). JT gravity is a simple yet rich two-dimensional dilaton gravity theory with a negative cosmological constant. It exhibits holographic features and serves as a canonical example of the AdS$_2$/CFT$_1$ correspondence. As a solvable model with a wealth of quantum structures, JT gravity has been extensively investigated in \cite{Saad:2019lba,Maldacena:2016upp,Engelsoy:2016xyb,Iliesiu:2019xuh,Iliesiu:2020zld}. Importantly for our purposes, JT gravity captures the dynamics of the near-horizon region of extremal Reissner–Nordström black holes (see \cite{fabbri2005modeling} for a review), as well as the causal diamond \cite{Banks:2021jwj, Gukov:2022oed}. This connection enables us to relate the entanglement entropy computed from the two-dimensional corner algebra to physically relevant four-dimensional spacetimes, thereby recovering the area law, which constitutes one of the main results of the manuscript, and a primer explicit application of the corner proposal. 

The structure of the paper is as follows. In the next section we briefly review the construction of the corner symmetry group in two dimensions performed in \cite{Ciambelli:2024qgi} and the procedure for gluing segments. Section \ref{reglu} is devoted to the representation theory of the QCS group and provides an explicit description of the gluing process in this context. In section \ref{entent} we turn to the central theme of our work: computing the entanglement entropy associated with two glued segments, first in the vacuum state and then in a coherent state. This analysis sheds light on both the quantum and semiclassical aspects of our system. In section \ref{sec:rényientropy}, we compute the Rényi entropy and, by identifying the Rényi index $w$ with the (dimensionless) inverse temperature, investigate the properties of the modular Hamiltonian and capacity of entanglement. Section \ref{sec:arealaw} is devoted to establishing the connection between our earlier results and four-dimensional spherically symmetric spacetimes dimensionally reduced to two dimensions. We find that, in the near-horizon geometry of the near-extremal Reissner--Nordström black hole, the entanglement entropy of a specific type of coherent states correctly reproduces the Bekenstein--Hawking area law.
 This is the result of the application of the corner proposal to $2$d JT gravity. In the final section \ref{genqu}, we construct the most general corner state as a combination of states belonging to different representations, generalizing the non-Abelian gauge theory results of  \cite{Donnelly:2014fua}. We show that this general state can be interpreted as a superposition of fuzzy area states sharply peaked around a classical value. We then conclude in section \ref{concl} recapitulating our findings and offering future directions. Technical details on the dimensional reduction of Noether charges are relegated to appendix \ref{appendix:dimred}. Similarly, the construction of the momentum map from the gravitational phase space to the coalgebra is exhiled to appendix \ref{appendix:momentummaps}. Finally, appendix \ref{appendix:relent} contains a digression on the relative entanglement entropy of coherent states.

\section{Quantum Corners}\label{sec:quantumcorners}

In this section, we review the concept of corners in gravitational theories, the essence of the corner proposal, the representations and gluing procedure.

\subsection{Corner Charges and Algebras}\label{coalg}

The understanding of the role of gravitational surface charges -- localized on co-dimension 2 surfaces -- first emerged in the work of Regge and Teitelboim~\cite{Regge:1974zd}. More recently, it has been understood that, in the presence of boundaries, certain diffeomorphisms acquire a non-vanishing Noether charge on the co-dimension 2 surface on which a hypersurface intersects the boundary. These surfaces are referred to as corners. These charge, which are now physical, obey an algebra called the algebra of corner symmetries \cite{Donnelly:2016auv}. The modern framework of understanding these charges and their algebra relies on the covariant phase space formalism \cite{Crnkovic:1986ex,Lee:1990nz,Iyer94,Wald:1999wa}. We give here a brief summary  of the formalism for the special case where symmetries are diffeomorphisms and refer to \cite{Harlow:2019yfa,Ciambelli:2022vot,Ciambelli:2023bmn,Assanioussi:2023jyq} for a more complete treatment.\par

The covariant phase space formalism associates a phase space structure to the space of solutions of a field theory. 
 Given a set of dynamical fields $\varphi^a$ defined on a manifold $M$ with a boundary and a Lagrangian top-form $L$ (possibly containing total derivatives), one can define a symplectic potential current $\theta$ via
\begin{equation}\label{eq:variationlagrangian}
    \delta L = E_a \delta \varphi^{a} + \dd \theta,
\end{equation}
where $E_a$ are the equation of motions and the index $a$ labels all the different fields of the theory. In the above, $\delta$ must be understood as an exterior derivative on field space. Thus the symplectic current is a $d-1$ form on spacetime and a one-form on field space.
Before proceeding, we clarify that, in this work, a theory is defined by specifying both a bulk Lagrangian and a boundary Lagrangian, rather than by treating the Lagrangian as defined only up to total derivative terms.

Integrating the current over a hypersurface $\Sigma$ yields the symplectic potential
\begin{equation}
    \Theta = \int_\Sigma \theta.
\end{equation}
The symplectic $2$-form is obtained by varying the above
\begin{equation}
    \Omega = \delta \Theta.
\end{equation}
We can realize any diffeomorphism $\xi \in \mathrm{T}M$ on field space by constructing the vector field ($\ldv{\xi}$ denotes the spacetime Lie derivative)
\begin{equation}\label{eq:fieldspacevectorfield}
    V_{\xi} = \int_M \dd^d x\, \ldv{\xi}\varphi^a \fdv{\varphi^a}.    
\end{equation}
 The contraction with the symplectic form can in general be written
\begin{equation}
I_{V_\xi} \Omega = -\delta H_{\xi} + F_{\xi},\label{IO}
\end{equation}
where $H_\xi$ is the Noether charge associated with the diffeomorphism and $F_\xi$ is the symplectic flux. Since the symmetries are diffeomorphisms, Noether second theorem implies that, on shell, the Noether charge is supported on the boundary
\begin{equation}
    H_{\xi} \hat = \int_{\partial \Sigma}Q_\xi.
\end{equation}

The presence of a symplectic flux spoils the integrability of the charge, that is, makes the right-hand side of equation \eqref{IO} not $\delta$-exact. This flux is associated with diffeomorphisms that move the boundary. In order to make any diffeomorphism integrable, we will work in the extended phase space formalism \cite{Ciambelli:2021nmv, Klinger:2023qna}, see also \cite{AliAhmad:2024vdw}. In that framework, the embedding of the boundary into the manifold is accounted for in the phase space by augmenting it with embedding maps, interpreted as edge modes, which restore the integrability of charges. The symplectic form is then extended by an additional corner term
\begin{equation}
    \Omega^{\mathrm{ext}} = \Omega + \int_{\partial\Sigma}\phi^*\qty(\iota_\chi \theta + \frac12 \iota_\chi \iota_\chi L),
\end{equation}
where $\phi$ is the corner embedding map and $\chi$ is a spacetime vector field and a field space one form defined by the embedding map 
\begin{equation}
    \chi^\mu = \delta \phi^\mu \circ \phi^{-1}.
\end{equation}
It can then be shown that, for any diffeomorphism, we obtain integrable Noether charges
\begin{equation}
I_{V_\xi}\Omega^{\mathrm{ext}} = \delta \qty(\int_{\partial\Sigma} \phi^*\qty(Q_\xi)),
\end{equation}
where the local Noether current is defined via
\begin{equation}\label{eq:noethercurrent}
    \dd Q_{\xi} = I_{V_\xi}\theta -\iota_{\xi}L.
\end{equation}

In gravitational theories, a subset of diffeomorphisms therefore acquire a physical integrable charge with support on the corner. 
The symplectic form can further be used to define the algebra of those charges
\begin{equation}
    I_{V_{\xi_1}} I_{V_{\xi_2}} \Omega^{\mathrm{ext}} =- \delta_{\xi_1} H_{\xi_2} \defeq \qty{H_{\xi_1},H_{\xi_2}},
\end{equation}
which represents the diffeomorphism algebra up to central extension
\begin{equation}
\qty{H_{\xi_1},H_{\xi_2}} = - H_{[\xi_1,\xi_2]} + c(\xi_1,\xi_2).
\end{equation}
This is the origin of the corner symmetry groups. Indeed, we have shown that, on the extended phase space, the algebra of charges furnish a projective representation of the algebra of symmetries at the corner of interest, which is simply dictated by the Lie algebra of vector fields.

Different formulations of gravity, even when describing the same underlying dynamics, can possess distinct corner symmetry groups. This naturally leads to the question: what is the maximal symmetry group that a gravitational theory can admit? It turns out that, as a purely kinematical statement, diffeomorphisms in a four-dimensional theory can at most realize the so-called universal corner algebra (UCA) whose associated group is \cite{Ciambelli:2021vnn}
\begin{equation}
    \mathrm{UCA} = \mathrm{Diff}(S) \ltimes \qty(\mathrm{GL}\qty(2,\RR)\ltimes \RR^2)^S,
\end{equation}
where $S$ denotes the corner, the general linear group is associated with linear transformations in the two-dimensional plane normal to the corner and $\mathbb{R}^2$ are the normal translations. The $S$ in the exponent emphasizes that there exists a copy of the subgroup $\mathrm{GL}\qty(2,\RR)\ltimes \RR^2$ at each point of the corner. The statement, then, is that the corner symmetries of any gravitational theory must form a subgroup of the universal corner symmetry group. For example, in the context of asymptotic symmetries of asymptotically flat spacetimes with Einstein-Hilbert gravity, one finds the so-called Weyl BMS group \cite{Freidel:2021fxf}
\begin{equation}
    \mathrm{BMSW} = \qty(\mathrm{Diff}\qty(S)\ltimes \RR^S) \ltimes \RR^S ,
\end{equation}
which is indeed a subgroup of the UCA where only the trace part (Weyl factor) of the general linear group is activated and only one super-translation survives. On the other hand, considering corners situated at finite distances for the same theory yields the extended corner symmetry group
\begin{equation}\label{eq:ecsgroup}
    \mathrm{ECS} = \mathrm{Diff}\qty(S)\ltimes \qty(\spl{2}\ltimes \RR^2)^S.
\end{equation}
The group includes the original corner symmetries identified in \cite{Donnelly:2016auv}, along with additional normal translations. As mentioned earlier, these translations can be interpreted as canonical charges through the extended phase space formalism \cite{Ciambelli:2021nmv}. While there is no systematic understanding of this observation, it appears that the trace part of the general linear group is associated with asymptotic symmetries, yet it is absent in the analysis of finite-distance corners. This is of course consistent with its interpretation in terms of a Weyl factor, as the latter is expected to correspond to dilatation of the corner. The ECS group is therefore the natural starting point in the description of local subsystems divided by a finite distance entangling corner.

In the context of the corner proposal, quantum gravity states are postulated to be described by the representation theory of the corner symmetry group. Although the full representation theory of the \(\mathrm{ECS}\) group is not known, the corner proposal can first be tested in the two-dimensional setting, the study of which was initiated in the authors' previous work \cite{Ciambelli:2024qgi}. Before briefly reviewing the initial results, we note that the formalism developed in this work can be interpreted as a per-point analysis of the higher-dimensional case. More precisely, the two-dimensional formalism captures the \( s \)-mode sector of the gravitational field, as we will clarify in section \ref{sec:arealaw} and appendix \ref{appendix:dimred}.

The two-dimensional extended corner symmetry group is given by
\begin{equation}\label{eq:twodimensionalecs}
    \mathrm{ECS} = \spl{2}\ltimes \RR^2.
\end{equation}
Since the associated quantum theory is defined on the projective representations of the above, we consider the maximally centrally extended version which is called the quantum corner symmetry group (QCS) \cite{Ciambelli:2024qgi,Varrin:2024sxe}
\begin{equation}
    \mathrm{QCS} = \reallywidetilde{\mathrm{SL}}\qty(2,\RR)\ltimes H_3,
\end{equation}
where the tilde denotes the universal cover and $H_3$ is the three-dimensional Heisenberg group. 

Let us now consider a corner describing the boundary of a subregion. The quantum state of this corner is given by a vector in the representation space of the QCS. Suppose we now have two corners with distinct states, \(\ket{\psi_L}\) and \(\ket{\psi_R}\), and we wish to glue them together to form the entangling surface between the two subregions. As mentioned in the introduction, it is a well known fact that the naïve tensor product state
\begin{equation}
    \ket{\tilde{\psi}}_G = \ket{\psi_L} \otimes \ket{\psi_R},
\end{equation}
 fails to accommodate the constraints of a gauge theory. The seminal work of Donnelly and Freidel \cite{Donnelly:2016auv}, from which the modern understanding of corner symmetries originates, introduces an entangling product to construct a gauge-invariant glued state. Their construction relies on the introduction of edge modes that restore gauge invariance by forcing the diffeomorphism corner charge to vanish.
In the present work, however, the charges are regarded as physical observables, whose possible values organize the states of the theory. We will therefore adopt the gluing procedure proposed in \cite{Ciambelli:2024qgi}, which simply states that two corners are identified if the values of the corner charges associated with the left and right subregions coincide. That is, given a maximally commuting subalgebra of the QCS, $\mathfrak{qcs}_{\mathrm{MC}} \subset \mathfrak{qcs}$, the glued Hilbert space is defined as
\begin{equation}
    \mathcal{H}_G = \qty{\ket{\psi_G}\in \mathcal{H}_L \otimes \mathcal{H}_R\mid Q_L \ket{\psi_G} = Q_R \ket{\psi_G}, \forall Q \in \mathfrak{qcs_\mathrm{MC}}},
\end{equation}
where the $L/R$ subscripts denote the generators of the algebra associated with the left/right subregion. The next section is devoted to the explicit construction of this gluing procedure.

\subsection{Representations and Gluing}\label{reglu}

We denote the basis of the the $\mathfrak{qcs}$ algebra by
$\left\{L_0,L_\pm,P_\pm,C\right\}$, where the $L$ operators form the $\mathfrak{sl}\left(2,\mathbb{R}\right)$ subalgebra
\begin{equation}
    [L_0,L_\pm] = \pm L_\pm, \qquad [L_-,L_+] = 2L_0,
\end{equation}
and the remaining elements generate the Heisenberg algebra
\begin{equation}\label{eq:heisenbergalgebra}
    [P_-,P_+] = C,
\end{equation}
where $C$ is a central element. The semi-direct product structure implies the following cross commutation relations
\begin{equation}
    [L_0,P_\pm] = \pm \frac12 P_\pm, \quad [L_\pm,P_\mp] = \mp P_\pm.
\end{equation}
The QCS has two Casimir operators: the central element $C$, and a cubic Casimir which is a generalization of the cubic Casimir in \cite{Ciambelli:2022cfr}. It reads
\begin{equation}\label{eq:qcscubiccasimir}
      \mathcal{C}_{\mathrm{QCS}} = C\qty(L_0( L_0 + \frac32)- L_- L_+) +\frac12\left(L_- P_+^2 + L_+ P_-^2 -2 L_0 P_- P_+\right).
\end{equation}

The construction of the unitary irreducible representations of the $\mathrm{QCS}$ were described in details in \cite{Varrin:2024sxe}, and related to the orbit method in \cite{Neri:2025fsh}. It turns out that the representations can be written as a tensor product of the standard $\reallywidetilde{\mathrm{SL}}\qty(2,\RR)$ representations with the representations of the Heisenberg group, where the special linear operators act on the formers through the metaplectic (or Weil) representation. In the present paper, we will focus on the positive discrete series, as the existence of a lowest weight state allows for a natural definition of the vacuum state. The discrete series representations can be written on the Hilbert space
\begin{equation}
\mathcal{H}^{(s)}_\mathrm{QCS} = \qty{\ket{n,k} \mid n,k \in \mathbb{N}},
\end{equation}
with the scalar product $\braket{n,k}{n',k'}
= \delta_{nn'}\delta_{kk'}$. The algebra generators then act as
\begin{Align}\label{discreteseriesrepresentation}
    C\ket{n,k} &=  c\ket{n,k},\\
    P_- \ket{n,k} &= \sqrt{ ck} \ket{n,k-1},\\
    P_+ \ket{n,k} &= \sqrt{c (k+1)} \ket{n,k+1},\\
    L_0 \ket{n,k} &= \qty(s +\frac{k}{2}+n+\frac{3}{4})\ket{n,k},\\
    L_- \ket{n,k} &= \sqrt{n (n+2 s)}\ket{n-1,k} + \frac12\sqrt{k(k-1)}\ket{n,k-2},\\
    L_+ \ket{n,k} &= \sqrt{(n+1) (2 s +n+1)}\ket{n+1,k}+\frac12\sqrt{(k+1)(k+2)}\ket{n,k+2},
\end{Align}
where $c\in \mathbb{R}$ is the scalar value of the central Casimir, and where $s \in \RR^+$ is related to the Casimir operator by
\begin{equation}\label{eq:casimirqcsrep}
\mathcal{C}_{\mathrm{QCS}}\ket{n,k} = c\qty(s^2-\frac{1}{16})\ket{n,k}.
\end{equation}

An interesting feature of the $\mathrm{QCS}$ representations is that the QCS cubic Casimir is equal to the value that the $\spl{2}$ cubic Casimir takes in its irreducible representations, multiplied by the central Casimir value. Since we are working in a two-dimensional setting and $\spl{2}$ is the conformal group in one dimension, one can take inspiration from the  AdS$_{2}$/CFT$_{1}$ correspondence to get a physical intuition behind the QCS Casimir. Indeed, in the AdS/CFT correspondence, the Casimir of the conformal algebra in the boundary theory is related to the Beltrami-Laplace operator in the bulk
\begin{equation}\label{boundaryasimirbulkenergy}
    s =  \sqrt{1/16 + (m L_{\mathrm{AdS}})^2}.
\end{equation}
where $m$ is the mass of a bulk field and $L_\mathrm{AdS}$ is the length scale of the $\mathrm{AdS}$ spacetime \cite{Witten:1998qj}. 

For large energy values with respect to the curvature associated with the length scale, the QCS Casimir can therefore be informally interpreted as the product of the typical energy of the bulk fields with a fundamental length scale of the spacetime
\begin{equation}
    \mathcal{C}_{\mathrm{QCS}} \sim \qty(E \alpha).
\end{equation}
This length scale will appear again in the following discussion of the gluing procedure. Since the $P_\pm$ generators can always be rescaled to set the proportionality constant $c=1$, we will adopt this choice from now on and restore the general value when relevant.

 In \cite{Ciambelli:2024qgi}, it was proposed that gluing two corners can be achieved by first taking the tensor product of the corner states and then imposing that the charge values at the corner for the left and right subsystems are equal. Equality of charges at the quantum level was taken to mean that the value of the maximally commuting subalgebra should coincide. This means that the task of gluing corners is reduced to finding the common eigenbasis for a maximal number of algebra operators. In order to do so, we will work in the following basis of Hermitian operators
\begin{Align}\label{Hermitianbasis}
    H &= \frac{1}{\alpha}\left(L_0 -\frac12 \qty( L_++L_-)\right),\\
    D &= \frac{i}{2}(L_+ - L_-),\\
    K &= \alpha\left(L_0 + \frac12(L_++L_-)\right),\\
    X &= \frac{1}{\sqrt{2 }}(P_++P_-),\\
    P &= \frac{i}{\sqrt{2 }}(P_+-P_-).
\end{Align}
The coefficient $\alpha$ has the dimension of length and is introduced to assign canonical dimensions to the conformal algebra. Similarly to \cite{deAlfaro:1976vlx}, $\alpha$ is an arbitrary scale that, once fixed, plays a fundamental role in the physics of the system, as for instance that of the $\mathrm{AdS}$ radius in $\mathrm{AdS}/\mathrm{CFT}$. The conformal mechanics states used here also appear in the study of the causal diamond in Minkowski spacetime \cite{Arzano:2020thh,Arzano:2021cjm,Arzano:2023pnf}. There, the parameter $\alpha$ is related to the size of the causal diamond. It is therefore natural to think of this scale as defining the typical length of our subsystem.  As we will see later, it is related to the Unruh temperature associated with a corner.

We also note that the Heisenberg algebra remains dimensionless. The cross commutator now become
\begin{equation}
    \qty[H,X] = -\frac{i}{\alpha} P, \quad [K,P] = i\alpha X,\quad \qty[H,P] = \qty[K,X] = 0.
\end{equation}
From the above equations, we can see that one choice of maximally commuting subalgebra is given by\footnote{In principle, the maximally commuting subalgebra should also contain the central generator $C$. Since our choice of normalization makes it the unit operator, we can safely leave it out in the present work.}
$\mathcal{C}_\mathrm{QCS},H,P$.
Due to the tensor product structure of the representation, we can consider the $\spl{2}$ and the Weil group generated by $X$, $P$ independently. The eigenstates of $P$ are simply given by 
\begin{equation}
    \ket{p}= \sum_{k=0}^\infty h_k(p) \ket{k} = e^{-p^2/2}\sum_{k=0}^\infty \qty(\frac{1}{\sqrt{\pi }2^k k!})^{\frac12} H_k\qty(p) \ket{k},
\end{equation}
where $H_k$ are the Hermite polynomials and the normalization coefficient of the wave function was chosen such that
\begin{equation}
   \braket{p}{p'} = \delta(p-p').
\end{equation}

For the $\spl{2}$ side, the eigenstates of $H$ are given by \cite{deAlfaro:1976vlx}
\begin{equation}\label{Ebasis}
    \ket{E} = \sum_{n=0}^\infty C_n(E) \ket{n},
\end{equation}
where
\begin{equation}
    C_n(E) = (2\alpha)^{s+\frac12} \left(\frac{\Gamma(n+1)}{\Gamma(n+2s+1)}\right)^\frac12 E^{s} e^{-\alpha E} L_n^{2s}(2 \alpha E),
\end{equation}
where $L_n^{2s}$ are the generalized Laguerre polynomials and where $E$ covers the positive real axis. The normalization coefficient was chosen such that
\begin{Align}
   \braket{E'}{E} &=  \sum_{n=0}^{\infty}C_n(E) C^*_n(E) = \delta(E'-E).
\end{Align}

The gluing procedure thus forces the left and right states to have the same values of $E$ and $p$. The inverse procedure, which will be referred to as the doubling map can be expressed as
\begin{equation}
    \ket{\psi,\phi} = \int \dd\Omega_{p,E}\,\psi(E)\phi(p)\ket{E,p} \longrightarrow \int \dd \Omega_{p,E} \, \psi(E)\phi(p) \ket{E,p}_L\otimes \ket{E,p}_R,
\end{equation}
where we defined the measure on the gluing basis
\begin{equation}
    \int \dd \Omega_{p,E} = \int_{-\infty}^\infty \dd p \int_0^\infty \dd E.
\end{equation}

The doubling map allows one to factorize the Hilbert space associated with a corner into its local subsystems components. We note that this operation resembles the factorization map in \cite{Jafferis:2019wkd}, but here, the use of symmetries uniquely fixes the degeneracy function appearing in the factorization.\par
While the representation theory of the $\mathrm{QCS}$ underpins the kinematics of the corner proposal, this map introduces additional information also derived from the symmetries. This extra structure connects to the dynamics of the original theory. Indeed, the procedure is built upon the value of the charges. Those are computed on-shell and know about the dynamics of the theory through Hamilton's equation. This interpretation is reminiscent of tensor networks, where, similarly, the factorization of the Hilbert space provides the additional structure missing from a simple quantum phase space without dynamics \cite{Cao:2016mst}.
\section{Entanglement Entropy}\label{entent}

In this section, we study the entanglement entropy of 2 important classes of states, the vacuum state and coherent states. While the former is constructed naturally form the lowest weight representation the latter is a combination of the already-known coherent states for the two algebra subsectors. We then analyze the behavior of the entropy for small and large values of the Casimir. We conclude this section with an overview of the Rényi entropy and related quantum informational properties of our states, such as the modular Hamiltonian and the related capacity of entanglement.

\subsection{Vacuum state}\label{vasta}
Given the above description of quantum states associated with a corner separating two subregions, we can ask what the entanglement entropy between these subsystems is.

We begin with the vacuum state, defined as the lowest weight state of the representation, that is
\begin{equation}
    \ket{\Omega} = \ket{n=0,k=0}.
\end{equation}
We can express this state in the gluing basis
\begin{Align}
    \ket{\Omega} &= \frac{\qty(2\alpha)^{s+\frac12}}{\pi^\frac14} \qty(\Gamma(2s+1))^{-\frac12}\int \dd \Omega_{p,E} \,E^{s} e^{-\alpha E}\, e^{\frac{-p^2}{2}} \ket{E,p}.
\end{Align}
Using the doubling map and tracing over the right degrees of freedom yields the reduced density matrix of the vacuum state
\begin{equation}\label{vacuumreduceddensityoperator}
    \rho^{\Omega}_L  = \frac{(2\alpha)^{2s+1}}{\sqrt{\pi}\,\Gamma(2s+1)}\int \dd \Omega_{p,E}\, E^{2s}e^{-2\alpha\left(E+\frac{p^2}{2\alpha}\right)} \ket{E,p}\bra{E,p}.   
\end{equation}
This density matrix represents a thermal state with temperature $T=2\alpha$, whose 
density of states is controlled by the parameter $s$. Its continuous energy levels are given by $\mathcal{E} = E + \frac{p^2}{2\alpha}$. 

\begin{figure}[H]
    \centering
    \includegraphics[width=0.6\linewidth]{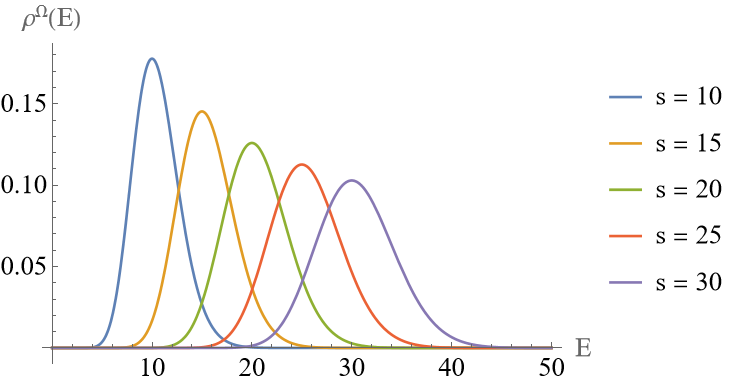}
    \caption{The distribution associated with the $\spl{2}$ part of the reduced density matrix of the vacuum state for unit temperature and different values of representation parameter $s$.}
    \label{fig:rhoE}
\end{figure}

One observes that the operator factorizes into independent distributions
\begin{Align}
    \rho_L^\Omega(E)=\frac{(2\alpha)^{2s+1}}{\Gamma(2s+1)}E^{2s}e^{-2\alpha E}\qquad\qquad 
    \rho_L^\Omega(p)=\frac{1}{\sqrt{\pi}}e^{-p^2}.
\end{Align}
The first of the above corresponds exactly to the vacuum state density operator of conformal quantum mechanics \cite{deAlfaro:1976vlx}, and is indeed the part related to $\spl{2}$. We plot this distribution for unit temperature and different values of $s$ in Fig.\ref{fig:rhoE}. 
The average energy of the distribution is given by\footnote{The expectation value of the operator $H$ is the energy since $L_0$ is the Hamiltonian and $\alpha \ev{H}_0 = \ev{L_0}_0$.}
\begin{equation}\label{groundstateenergy}
    \ev{H}_0 = \frac{s+\frac12}{\alpha}. 
\end{equation}
In the large $s$ limit, the above equation is reminiscent of the holographic result \eqref{boundaryasimirbulkenergy}. Moreover, as already observed, the partition function in \eqref{vacuumreduceddensityoperator} indicates a thermal behavior of the vacuum once restricted to the subregion. This suggests the following interpretation. As mentioned in \cite{Ciambelli:2024qgi}, since these states are gravitational states, the vacuum state should be interpreted as the absence of geometry.\footnote{This is clearly stated in the theory of embadons explored in \cite{Ciambelli:2024swv}, in which the area of a corner is promoted to a quantum operator, and thus its spectrum becomes only semi-positive definite. Then, the vacuum state is associated with the zero eigenvalue of the area operator, interpreted as the absence of geometry, rather than as the geometry being flat. Indeed, the former interpretation pertains to quantum geometric states, while the latter would be the case for an operator associated to matter (or perturbative gravitational radiation) on a classical geometric background.} Nonetheless, such a claim is made at the level of the global pure state, before partitioning it into subsystems. In fact, when an observer is restricted to a subsystem and does not have access to the information on the other side of the entangling corner, she traces out the degrees of freedom in the complementary inaccessible region, inevitably resulting in her experiencing thermal physics \cite{Gibbons:1977mu}. This phenomenon, originally discovered by Unruh in the framework of quantum field theory \cite{Unruh:1976db}, is a robust algebraic feature \cite{Bisognano:1976za} that we have discovered here applies also to our quantum geometric description. 

Before moving to the entanglement entropy, we note that the argument of the exponential in \eqref{vacuumreduceddensityoperator} is exactly the eigenvalue of the full operator $H$
\begin{equation}
    H\ket{E,p} = \left(E + \frac{p^2}{2\alpha}\right) \ket{E,p}.
\end{equation}
This allows us to write the reduced density operator in a basis-independent way
\begin{equation}
    \rho_L^\Omega = \frac{(2\alpha)^{2s+1}}{\Gamma(2s+1)} \left(H-\frac{P}{2\alpha}\right)^{2s} e^{-\alpha H}.
\end{equation}
The presence of the $P$ operator differs from the vacuum density operator of \cite{deAlfaro:1976vlx}, illustrating the presence of the additional Heisenberg structure (that is, the translations) in our system. It is interesting to observe that we recover the de Alfaro-Fubini-Furlan theory  of \cite{deAlfaro:1976vlx} in the high temperature limit. We further remark that the modular Hamiltonian of the system defined by the entangling corner can be deduced from the equation above. In the $s = 0$ case, for example, it takes the simple form $H_{\mathrm{mod}} = \alpha H$. It would therefore be of great interest to study the Rindler physics associated with such a Hamiltonian, as this could potentially offer insights into the spacetime interpretation of the underlying dynamics, making the above thermal discussion more rigorous.

We now turn our attention to the entanglement entropy.
In order to compute the von-Neumann entropy of the vacuum state, we start by rewriting the density operator in dimensionless variables. We define $\tilde{E} = \alpha E$ and write
\begin{equation}\label{vacredden}
    \rho_L^\Omega = \frac{2^{2s+1}}{\sqrt{\pi}\Gamma(2s+1)} \int \dd p\dd \tilde{E} \, \tilde{E}^{2s}e^{-\left(2\tilde{E} + p^2\right)} \ket{E,p}\bra{E,p}.
\end{equation}
The vacuum entanglement entropy associated with the subsystem is then
\begin{equation}\label{vacententr}
    S^\Omega \equiv -\mathrm{Tr}\left(\rho_L^\Omega \ln \rho_L^\Omega \right) = (2s + 1)+ \ln(\frac{\Gamma(2s+1)}{2}) -2s \psi^{(0)}(2s+1) + \lambda,
\end{equation}
where $\psi^{(0)}$ is the digamma function and the constant $\lambda$ comes from the $p$ integral
\begin{equation}
    \lambda = \frac12(1 + \ln(\pi)).
\end{equation}
Note that, since the global state was pure,  we have dropped the $L$ index on the entropy since the right subregion yields the same result. 

The vacuum entropy is therefore given by a monotonously increasing function of the representation parameter $s$. For small values of the representation parameter $s$, the entropy is linear
\begin{equation}
    S^\Omega \sim 2s, \quad s\ll1.     
\end{equation}
As $s$ grows, the entanglement entropy transition into a logarithmic behavior
\begin{equation}
    S^\Omega \sim \frac12 \ln(s), \quad  s\gg1.
\end{equation}
Since the effective dimension of the system associated with the vacuum state is given by $d_{\mathrm{eff}} \sim e^{S^\Omega}$, this result suggests that, for large \( s \), the representation parameter can be interpreted as the effective number of degrees of freedom of the system. We will revisit this idea later when dealing with the general corner states.

Although the vacuum state is the natural starting point for the calculation of entanglement entropy, we argued above that it should be associated with the absence of geometry. Moreover, it does not saturate the uncertainty relation. As such, it is not well-suited to describe the appearance of classical geometry from a semiclassical limit. This motivates us to study coherent states, and their entanglement entropy.

\subsection{Coherent states}\label{costa}

Loosely speaking, coherent states are quantum states that look the most classical. More precisely, one of the defining feature of coherent states is that they saturate the uncertainty principle. How do we construct coherent states for the QCS? Since the representation factorizes into the special linear and the Heisenberg representations, it is natural to start by looking at the two cases separately.

Let us first focus on the $\spl{2}$ part. Here, coherent states were found by Perelomov \cite{APerelomov_1977}. These states are parametrized by a complex number $\zeta \in \mathbb{C}$ with $|\zeta| < 1$, and are constructed by applying a displacement operator to the lowest-weight state
\begin{equation}\label{eq:displacementoperator}
    \ket{\zeta} = e^{c_\zeta L_+ - \bar{c}_\zeta L_-}\ket{n=0} \defeq \mathcal{S}(c_\zeta)\ket{n=0},
\end{equation}
where, denoting $\zeta = re^{i\theta}$
\begin{equation}
    c_\zeta =\tanh^{-1}(r)e^{i\theta}.
\end{equation}
They can be expressed in the discrete basis as
\begin{equation}
    \ket{\zeta} = (1-\abs{\zeta}^2)^{s+\frac12} \sum_{n=0}^\infty \left(\frac{\Gamma(n+2s+1)}{\Gamma(n+1)\Gamma(2s+1)}\right)^\frac12 \zeta^n \ket{n}.
\end{equation}

For the gluing procedure, we  need to express these states in the $\ket{E}$ basis. This is done using equation \eqref{Ebasis} and the generating function for Laguerre polynomials
\begin{Align}\label{perelomovinEbasis}
    \ket{\zeta} &= \frac{\qty(2\alpha)^{s+\frac12} (1-\abs{\zeta}^2)^{s+\frac12}}{\sqrt{\Gamma(2s+1)}} \int_0^\infty \dd E\,E^{s} e^{-\alpha E}\sum_{n=0}^\infty \zeta^n L_n^{2s}(2\alpha E)\ket{E}\\
    &=  \frac{\qty(2\alpha)^{s+\frac12}(1-\abs{\zeta}^2)^{s+\frac12}}{\sqrt{\Gamma(2s+1)}(1-\zeta)^{s+\frac12}} \int_0^\infty \dd E\, E^{s} \, e^{-\alpha E\left(\frac{1+\zeta }{1-\zeta }\right)}\ket{E}.
\end{Align}

The Perelomov coherent states saturate the Schrödinger-Robertson uncertainty relation \cite{Trifonov:2000su}. For any pair of operators $X,Y \in \mathfrak{sl}\qty(2,\RR)$, defining
\begin{equation}
    \Delta(XY) \defeq \frac12 \expval{XY + YX}- \expval{X}\expval{Y},
\end{equation}
the Schrödinger-Robertson uncertainty relation is
\begin{equation}
    \qty(\Delta X)^2\qty(\Delta Y)^2\geq \frac14 \abs{\expval{[X,Y]}}^2 + \qty(\Delta(XY))^2.
\end{equation}
These states are therefore the most classically-looking states in the special linear representations.

Next, we study the Heisenberg part, where we can use the well-known Glauber coherent states parametrized by a complex number $\gamma\in\mathbb{C}$ and defined by displacing the vacuum
\begin{equation}
    \ket{\gamma} = e^{\gamma P_+-\bar{\gamma}P_-}\ket{k=0} \defeq \mathcal{D}(\gamma)\ket{k=0},
\end{equation}
where we defined the displacement operator $\mathcal{D}(\gamma)$.
These states famously saturate the Heisenberg uncertainty principle. We further note that, because of the way the $\spl{2}$ generators act on the Heisenberg representations \eqref{discreteseriesrepresentation}, the action of the Perelomov displacement operator \eqref{eq:displacementoperator} on the Heisenberg coherent states creates a squeezed coherent state
\begin{equation}
    \ket{\gamma_{\zeta}} = \mathcal{S}(c_\zeta)\ket{\gamma}.
\end{equation}
For $\theta = 0$ and $\theta = \pi$, these states also saturate the Heisenberg uncertainty relation. In the following, we focus on the states with $\theta= \pi$. 

Putting things together, we propose that the QCS coherent states are therefore given by 
\begin{equation}\label{eq:generalqcscoherentstate}
    \ket{\zeta,\gamma_\zeta} = \mathcal{S}(c_\zeta)\mathcal{D}(\gamma)\ket{\Omega}.
\end{equation}
They can be expressed in the $\ket{E,p}$ basis as 
\begin{equation}\label{cost}
    \ket{\zeta,\gamma_\zeta} = \intpE{p}{E} \psi_\zeta(E)\phi_{\zeta,\gamma}(p) \ket{E,p},
\end{equation}
where $\psi_\zeta(E)$ is given in equation \eqref{perelomovinEbasis} and $\phi_{\zeta,\gamma}(p)$ is the wave function of the Heisenberg squeezed coherent state in the momentum basis \cite{Schumaker:1986tlu,gonzalez2021}
\begin{Align}\label{squeezedcoherentinpbasis}
    \phi_{\zeta,\gamma}(p) = \braket{p}{\gamma_\zeta} =\frac{e^{\frac{r}{2}}}{\pi^\frac14} e^{-\frac{1}{2} e^{2 r} (p-\sqrt{2}\Im(\gamma))^2}e^{-i p
   \sqrt{2}\Re(\gamma)+i \Im(\gamma)
   \Re(\gamma)}.
\end{Align}
 The states \eqref{eq:generalqcscoherentstate} are important in the corner proposal, as they correspond to quantum geometric configurations that admit a semiclassical limit, as we will see in Section~\ref{sec:arealaw}.

We can now move on to the entanglement entropy. Because of the gluing and reducing procedures, the crucial quantities are the modulus square of the above wave functions
\begin{Align}\label{modulussquarewavefunctioncoherentsate}
     \abs{ \psi_\zeta(E))}^2 &= \frac{(2\alpha)^{2s+1}}{\Gamma(2s+1)}\left(\frac{1-r}{1+r}\right)^{2s+1} E^{2s}e^{-2 \alpha E \qty(\frac{1-r}{1+r})},\\
     \abs{\phi_{\zeta,\gamma}}^2 &= \frac{e^{r}}{\pi^\frac12}e^{-e^{2r}(p-\sqrt{2}\Im(\gamma))^2}.
\end{Align}
The reduced density matrix then takes the form
\begin{equation}
    \rho^{\zeta,\gamma}_L = \left(\frac{e^{r}}{\sqrt{\pi}}\right)\frac{(2\alpha)^{2s+1}}{\Gamma(2s+1)}\qty(\frac{1-r}{1+r})^{2s+1} \int \dd \Omega_{p,E}\,E^{2s}e^{-2\alpha E \,\qty(\frac{1-r}{1+r})-e^{2r}\left(p - \sqrt{2}\Im(\gamma)\right)^2} \ket{E,p}\bra{E,p}.
\end{equation}
Comparing the above to the vacuum state reduced density operator \eqref{vacuumreduceddensityoperator}, we see that the factors $\frac{1-r}{1+r}$ and  $e^{2r}$ act as temperature multipliers, while the imaginary part of $\gamma$ acts as shift in the Weil energy. The vacuum case is recovered for $r=\gamma=0$.

From the reduced density matrix, we can go to dimensionless variables and calculate the entanglement entropy, finding
\begin{equation}\label{eq.entanglemententropycoherentstate}
    S^{(\zeta,\gamma)} = S^\Omega + 2\tanh^{-1}(r)-r,
\end{equation}
where we recall that $S^\Omega$ is the vacuum entanglement entropy given in equation \eqref{vacententr}. The $r$ parameter controls how much the entanglement entropy of the coherent state deviates from that of the vacuum state. This behavior is expected, as $\frac{1-r}{1+r}$ and $e^{2r}$ effectively act as temperature multipliers. In contrast, the $\gamma$ parameter, which simply corresponds to a constant energy shift, does not affect the entanglement entropy. Interestingly, for Glauber coherent states, this result is well known in the context of quantum field theory and gravity in two dimensions, where it is related to the conformal flatness of two-dimensional geometries \cite{Fiola:1994ir,Katsinis:2022fxu,Das:2005ah,Varadarajan:2016kei}.

Since $r$ takes values in $[0,1[$, the entanglement entropy of the coherent state can be considerably larger than the one of the vacuum state. In particular, it is interesting to consider the family of states
\begin{equation}\label{eq:classicalstate}
    \ket{\psi_{cl}(s)} = e^{i s D}\ket{\Omega} = \ket{\zeta = -\tanh(s)}\otimes S(-s)\ket{k=0},
\end{equation}
corresponding to a Perelomov coherent state on the $\spl{2}$ side and a squeezed state with squeezing parameter $-s$ on the Weil side. In the application of this formalism to near-extremal black holes in section \ref{sec:arealaw}, these states can be identified with classical geometries, in the sense that they exhibit the correct classical limit, which is why we refer to them as classical states. Their entanglement entropy is given by 
\begin{equation}
    S^{\psi_{cl}} = S^\Omega+2s-\tanh(s).
\end{equation}
Conversely to the vacuum state, the state $\psi_{cl}$  has a linear dominant behavior in both $s$-limits
\begin{Align}\label{eq:asymptoticentropyclassicalstate}
    S^{\psi_{cl}} &\sim 2s + \frac12 \ln(s), \quad s\gg1,\\
    S^{\psi_{cl}} &\sim 3s,\quad s\ll1.
\end{Align}

We conclude this subsection noticing that one can use the classical understanding of the $\mathfrak{sl}(2,\mathbb{R})$ operators as generating boosts in the corner normal plane to familiarize with these states. Indeed, using the defining representation of the corner symmetry algebra \cite{Ciambelli:2022vot,Varrin:2024sxe} and picking spacetime coordinates $(u,\rho)$,\footnote{In particular, the corner is located at $u=0$ and $\rho=0$.} the classical version of the operator $D$ can be written as
\begin{equation}
    D = \frac12 \qty(u \partial_\rho + \rho \partial_{u}).
\end{equation}
In particular, in the basis $D,H,K$ of $\spl{2}$ it plays the role of $L_0$, and as such it corresponds to a boost along the spatial coordinate. This provides a physical interpretation of the coherent states \eqref{eq:classicalstate}, as those generated by the normal plane boost acting on the vacuum. Fascinatingly, this links the classical interpretation with the quantum description above, and provides a rationale for its thermal behavior once restricted to a subregion. Indeed, boosting a corner can be interpreted as moving along an accelerated trajectory in the normal plane, reminiscently of the Unruh effect.

We now turn our attention to the study of quantum informational properties of our states. For the coherent states of this section, we relegate to appendix \ref{appendix:relent} a digression on their relative entropy and its perturbations.

\subsection{R\'enyi entropy}\label{sec:rényientropy}

The R\'enyi entropy is a one-parameter generalization of the von Neumann entropy defined as
\begin{align}
    H_w = \frac{1}{1-w} \ln\left[\mathrm{Tr}\rho^w\right]
\end{align}
where $w\in\mathbb{R}^+$. For various values of $w$ the Rényi entropies capture more detailed information about entanglement spectra. Moreover, it is a useful computational tool to perform the replica trick in quantum field theory. Here, we are interested in exploring the quantum informational properties of our states.

In the case of the vacuum reduced density operator \eqref{vacredden}, one has
\begin{equation}\label{Renvacredden}
   \mathrm{Tr}\left( \qty(\rho_L^\Omega)^w\right) = \left(\frac{2^{2s+1}}{\sqrt{\pi}\Gamma(2s+1)} \right)^w\int_{-\infty}^\infty \dd p\int_{0}^\infty\dd \tilde{E} \, \tilde{E}^{2ws}e^{-w\left(2\tilde{E} + p^2\right)} = \left(\frac{2^{2s+1}}{\sqrt{\pi}\Gamma(2s+1)} \right)^w\, \sqrt{\frac\pi w}\, \frac{\Gamma(2ws +1)}{(2w)^{2ws +1}}
\end{equation}
and the R\'enyi entropy is
\begin{align}\label{Renent}
    (H_w)_L^\Omega = \frac{1}{1-w} \left[ (w-1)\left(\ln2 -\frac12\ln\pi\right) -\left(2ws+ \frac32\right)\ln w + \ln \Gamma(2ws +1) -  w\ln\Gamma(2s+1)
    \right]  
    \end{align}
Consistency is checked by taking the $w\rightarrow 1$ limit of $(H_w)_L^\Omega$ and ensuring it reproduces the entanglement entropy we computed above, \eqref{vacententr}.
We find
\begin{align}
    \lim_{w\rightarrow1} (H_w)_L^\Omega = \frac12\qty(\ln\pi +1)+\left(2s+ 1\right) - 2s\,\psi^{(0)}(2s+1) + \ln(\frac{\Gamma(2s+1)}{2}) = S^\Omega,
\end{align}
matching our previous result.

There are two interesting limits one can take, $w\to \infty$ and $w\to 0$. The former gives information on the largest eigenvalue of the density matrix, while the latter gives (for a finite-dimensional system) the logarithm of  its rank. In the limit $w \rightarrow\infty$ the Renyi entropy $(H_w)_L^\Omega$ becomes
\begin{align}
 \lim_{w \rightarrow\infty}(H_w)_L^\Omega &= -\lim_{w \rightarrow\infty} \frac1w\left(
  w\left(\ln2 -\frac12\ln\pi\right) -2ws\ln w + 2ws\ln 2ws -2ws -w\ln\Gamma(2s+1) \right)\nonumber\\
  &=  \frac12\ln\pi - 2s(\ln 2s -1) + \ln\frac{\Gamma(2s+1)}{2},
\end{align}
while at the limit $w \rightarrow 0$ diverges like $-3/2 \ln w$. This divergence is the sign that the density matrix has infinite rank, consistently with its continuous spectra.

For coherent states we can calculate
\begin{Align}\label{trrhowcoherent}
\Tr(\qty(\rho^{\zeta,\gamma})^w)&= \qty(\frac{e^{r}}{\sqrt{\pi}})^w\qty(\frac{2^{2s+1}}{\Gamma(2s+1)})^w \qty(\frac{1-r}{1+r})^{2sw+w}\int \dd \tilde{E} \dd p\, \tilde{E}^{2sw}e^{-2w\tilde{E}\qty(\frac{1-r}{1+r})-we^{2r}(p^2)},
\end{Align}
such that
\begin{equation}
    H_w^{\zeta,\gamma} = \left(\frac12 \ln(\pi) + \ln(\frac{\Gamma(2s+1)}{2}) + 2 \tanh^{-1}(r) -r\right)- \frac{\ln(w)}{1-w}\qty(2sw + \frac32) + \frac{1}{1-w}\ln(\frac{\Gamma(1+2sw)}{\Gamma(1+2s)}).
\end{equation}

One can check that the $w\rightarrow 1$ limit reproduces the Von-Neumann entropy
\begin{equation}
    \lim_{w\rightarrow 1} \qty(H_w^{\zeta,\gamma}) = S^\Omega +2 \tanh^{-1}(r)- r = S^{(\zeta,\gamma)}.
\end{equation}
Moreover, the $w \rightarrow \infty$ limit gives
\begin{equation}
    \lim_{w\rightarrow \infty}\qty(H^{\zeta,\gamma}_{w}) = \frac12 \ln \pi + \ln(\frac{\Gamma(2s+1)}{2}) + 2 \tanh^{-1}(r) -r -2s(\ln(2s)-1),
\end{equation}
while the $w\rightarrow 0$ limit diverges again as $-\frac32 \ln w$.

Recently, a new interpretation for the R\'enyi index $w$ has been proposed \cite{Baez:2011upp,Nakaguchi:2016zqi,DeBoer:2018kvc}, as the inverse temperature $\beta$ associated to the state.\footnote{Strictly speaking, since $w$ is dimensionless this is the inverse ration of the temperature $T$ divided by the reference temperature $T_0$.} Therefore we can interpret the limits $w \rightarrow\infty$ and $w \rightarrow0$ as low and high temperature limit, respectively. The finiteness of the R\'enyi entropy for $w \rightarrow \infty$ can be interpret as finiteness of the entropy at low temperatures. The standard thermal states share this behavior. On the other hand, the divergence at $w\rightarrow 0$ signals the proliferation of the number of states accessible at high temperature. It would be interesting to understand this behavior better.

Of particular interest is the capacity of entanglement, defined as 
\begin{align}
    C(\rho) =\beta^2 \left( \langle K^2\rangle - \langle K\rangle^2 \right),
\end{align}
where $\beta$ denotes the (dimensionless) inverse temperature, which below we will identify with the R\'enyi parameter $w$. The capacity of entanglement is proportional to the variance (the second cumulant) of the modular Hamiltonian $K$ associated with the reduced density matrix $\rho$.

Let us push the analogy between the Rényi index $w$ and the inverse temperature $\beta$ to its limits and, following \cite{DeBoer:2018kvc}, interpret the Rényi partition function as a thermodynamic partition function with (inverse) temperature $\beta$. 
\begin{align}\label{thermalpartf}
    Z_\rho (\beta) = \mathrm{Tr\,}\left( \rho^\beta \right)=\mathrm{Tr\,}\left( e^{-\beta K}\right)\,.
\end{align}
This is a partition function corresponding to a system of some fundamental degrees of freedom, whose thermodynamical properties we want to investigate.

From \eqref{thermalpartf} we can compute free energy
\begin{align}\label{thermalfreee}
    F_\rho(\beta) = - \frac1\beta\, \ln Z_\rho(\beta)\,,
\end{align}
and thermodynamical entropy
\begin{align}\label{thermalent}
 S_\rho(\beta)    = \beta^2 \frac{\partial F_\rho(\beta)}{\partial \beta} =  \left(\beta \partial_\beta -1\right) (\beta F_\rho(\beta) ).
\end{align}
This equation is equivalent to the first law of thermodynamics. Indeed, for $\beta = T_0/T$
\begin{align}
    dF_\rho(\beta) = \frac1{\beta^2} \,   S_\rho(\beta) = - S_\rho(\beta) d\left(\frac1{\beta}\right) = -S \frac{dT}{T_0}.
\end{align}
On the other hand \eqref{thermalent} can be rewritten as
\begin{align}\label{thermalent1}
 S_\rho(\beta)    =  \beta \left(\langle K\rangle_\beta -F_\rho(\beta) \right) 
\end{align}
which relates the entropy with the free energy  and the expectation value of the modular Hamiltonian. It is worth noticing the useful identities
\begin{align}
   \frac{\partial }{\partial \beta}\, (\beta F_\rho(\beta))&=  \langle K\rangle_\beta\\
    - \frac{\partial^2 }{\partial \beta^2}\, (\beta F_\rho(\beta))&= \langle (\Delta K)^2 \rangle_\beta \equiv \langle  K^2 \rangle_\beta  -  \langle  K \rangle^2_\beta.
\end{align}
Finally the heat capacity, called the entanglement capacity in \cite{Nakaguchi:2016zqi, DeBoer:2018kvc}, is
\begin{align}\label{thermalheatcap}
    C_\rho(\beta) =-\beta^2 \frac{\partial}{\partial\beta}\, \langle  K \rangle_\beta=\beta^2 \left(\langle K^2\rangle - \langle K\rangle^2\right)
\end{align}

After this brief review of the thermodynamic derivation of the entanglement capacity, let us consider these quantities in the specific case of the reduced vacuum density matrix. The partition function \eqref{thermalpartf} equals (see \eqref{Renvacredden})
\begin{align}\label{thpartvac}
    Z_\rho^\Omega (\beta) =  \left(\frac{2^{2s+1}}{\sqrt{\pi}\Gamma(2s+1)} \right)^\beta\, \sqrt{\frac\pi \beta}\, \frac{\Gamma(2\beta s +1)}{(2\beta)^{2\beta s +1}}.
\end{align}
The free energy is then
\begin{equation}
    F^\Omega_\rho(\beta) \equiv -\frac1\beta\, Z^\Omega_\rho(\beta) =\ln(\frac{\sqrt{\pi}\Gamma(2s+1)}{2^{2s+1}}) + \frac{1}{2\beta}\ln(\frac{\beta}{\pi}) - \frac{1}{\beta}\ln(\Gamma(2\beta s +1))+ \frac{2\beta s + 1}{\beta}\ln(2\beta),
\end{equation}
while the thermal entropy is 
\begin{align}\label{thermalentvac}
    S^\Omega_\rho(\beta) \equiv \beta^2 \frac{\partial F^\Omega_\rho(\beta)}{\partial\beta}= \frac{3+\ln\pi -2 \ln 2}{2} -\frac32\,\ln\beta+ 2\beta s(1- \psi^{(0)}(2\beta s +1)) + \ln\Gamma(2\beta s+1)
\end{align}
The expectation value of the modular Hamiltonian is
\begin{align}
   \langle K^\Omega\rangle_\beta  \equiv  \frac1\beta\left( S^\Omega_\rho(\beta) +\beta F^\Omega_\rho(\beta)\right)&=\frac3{2\beta} + 2s \qty(\ln\beta-\psi^{(0)}(2\beta s +1)+1) + \ln(\frac{\sqrt{\pi}\Gamma(2s+1)}{2}).
\end{align}
Finally, the capacity of entanglement is given by
\begin{align}
    \beta^2\left( \langle K^\Omega{}^2\rangle - \langle K^\Omega\rangle^2 \right)= -\beta^2\, \frac{\partial}{\partial\beta}\, \langle  K^\Omega \rangle_\beta  =\frac3{2} -2s\beta + (2s\beta)^2 \psi^{(1)}(2s\beta+1)\label{DeltaK}
\end{align}
where $\psi^{(1)}$ is the trigamma function.

In the small $\beta$ (high-temperature) limit, the entropy \eqref{thermalentvac} diverges logarithmically, while the capacity of entanglement remains constant. In the opposite limit of large $\beta$ (low temperature), the capacity of entanglement again remains constant, whereas the entropy becomes negative. The low temperature behavior is unphysical, as it contradicts the third law of thermodynamics. This suggests that at sufficiently low temperatures, the description of the system using the partition function \eqref{Renvacredden} breaks down, indicating that this partition function fails to capture essential physical information about the system’s behavior in this regime. One may speculate that in the low temperature regime the thermodynamic description proposed here breaks down, in favor of a more microscopic framework, suitable to capture this infinite-dimensional system with massless excitations.\footnote{The partition function \eqref{thermalpartf} is a thermal partition function, that could be equal to a more fundamental path integral partition function in some regimes.} Indeed, this situation is analogous to the case of the ideal gas at low temperatures, where the entropy also diverges in a similar manner. In the ideal gas case, the pathology is well understood: at low temperatures, quantum effects become significant, and the classical ideal gas partition function must be replaced with a quantum version that correctly describes the low-temperature physics. It would be interesting to similarly understand the low-temperature behavior of our system.

We now look at the coherent states whose partition function is given by (see equation \eqref{trrhowcoherent})
\begin{Align}
   Z^{(\zeta,\gamma)}(\beta) =  \sqrt{\frac{\pi}{\beta}} e^{r(\beta-1)}
 \left(\frac{1-r}{1+r}\right)^{\beta-1} \left(\frac{2^{2
   s+1}}{\sqrt{\pi}\Gamma (2 s+1)}\right)^\beta \frac{\Gamma (2 s \beta+1)}{(2\beta)^{2s\beta+1}}= \left(e^{r}\qty(\frac{1-r}{1+r})\right)^{\beta-1} Z^{\Omega}(\beta).
\end{Align}
The free energy is then
\begin{Align}
    F^{(\zeta,\gamma)}(\beta) &= F^{\Omega}(\beta) +\frac{1-\beta}{\beta}\qty(r -2\tanh^{-1}(r)),
\end{Align}
and the thermal entropy reads
\begin{equation}
    S^{(\zeta,\gamma)}(\beta) = S^{\Omega}(\beta) - r + 2 \tanh^{-1}(r).
\end{equation}
As expected, we find the same relation between the coherent states and the vacuum as in the case of the entanglement entropy~\eqref{eq.entanglemententropycoherentstate}.
Finally the expectation value of the modular Hamiltonian gives
\begin{equation}
  \expval{K}_\beta=2s \qty(\ln(\beta)-\psi^{(0)}(2\beta s +1)+1) + \ln(\frac{\sqrt{\pi}\Gamma(2s+1)}{2}) + 2\tanh^{-1}(r) -r.
\end{equation}
This implies that the fluctuations are the same for the vacuum or the coherent states. While coherent states are more suitable to describe semiclassical geometries, we have argue that the vacuum state is a purely quantum-geometric state of the system. That they experience the same fluctuations in their modular Hamiltonian is an interesting and unexpected feature. This means that by studying the semiclassical states and their fluctuations, we can infer some properties of the quantum gravity vacuum state.\footnote{In a subtle way, this is reminiscent of a UV/IR mixing effect, in which  IR physical predictions have effects on the UV sides of the theory.} 

We conclude remarking that there exist values of $s$ for which the fluctuations of these states are controlled by the modular Hamiltonian itself, that is, $\langle \Delta K^2\rangle = \langle K\rangle$, as advocated in \cite{Verlinde:2019xfb, Verlinde:2019ade, Verlinde:2022hhs, Ciambelli:2025flo,Fransen:2025npa}. Consider the vacuum state. From \eqref{DeltaK}, the fluctuation of the modular Hamiltonian is\footnote{We recall that $\beta$, being equal to the replica parameter, is dimensionless here.}
\beq
\langle \Delta K^2\rangle =\langle K^2\rangle - \langle K\rangle^2=\frac3{2\beta^2} -\frac{2s}{\beta} + (2s)^2 \psi^{(1)}(2s\beta+1),
\eeq
whereas the expectation value of the modular Hamiltonian is
\beq
\langle K\rangle  =\frac3{2\beta} + 2s \qty(\ln\beta-\psi^{(0)}(2\beta s +1)+1) + \ln(\frac{\sqrt{\pi}\Gamma(2s+1)}{2}).
\eeq
Now consider the limit $\beta\to 1$. Then, one can demonstrate that these two expressions are exactly equal whenever $s \approx 0.0326$, which is positive and thus allowed. Other states of this form can be obtained tuning $\beta$ and $s$ accordingly.\footnote{This argument can be applied to coherent states as well, thanks to the fact that their modular fluctuation is exactly equal to the vacuum one.} Therefore, we see that there exist states in our representations that have modular fluctuations proportional to the expectation value of the modular Hamiltonian. A more systematic analysis of this feature is part of our agenda.

\section{Application to Near-Extremal Black Holes}\label{sec:arealaw}
The first step in connecting the abstract formalism presented above to physical predictions is to verify that it reduces to known results in certain well-defined limits. In particular, a key consistency requirement for any candidate theory of quantum gravity is that the entanglement entropy exhibits an area law in the semiclassical regime.
 Although the formalism developed above originates from the corner symmetry group of two-dimensional gravity, it is also well suited to describing certain spherically symmetric spacetimes in higher dimension. This follows from the well-known fact that spherically symmetric geometries give rise, via dimensional reduction, to a two-dimensional dilaton gravity theory \cite{berger_1972,Solodukhin:1998tc,Grumiller:2002nm,Carlip:2017eud}. In particular, it has been shown in \cite{Banks:2021jwj,Gukov:2022oed} that the near-horizon geometry of a causal diamond in Minkowski space is described by JT gravity. The symmetries and charges of dimensionally reduced spherically-symmetric spacetimes were also recently discussed in \cite{BenAchour:2023dgj}.

First of all, let us remark on the corner symmetries perspective of this dimensional reduction.
 As mentioned earlier, the corner symmetry group of four-dimensional Einstein–Hilbert theory for a finite-distance corner \(S\) is given by the ECS \eqref{eq:ecsgroup}, whose structure entails a copy of the two-dimensional ECS group at each point of \(S\). Thus, the special linear and normal translation charges of the four-dimensional theory admit a spherical-harmonic expansion. In a spherically symmetric spacetime, the charges— which depend only on metric components and their derivatives—are independent of the angular position on the sphere, so only the zero mode in the spherical-harmonic expansion contributes.
 In other words, only the \(s\)-wave components of the charges matter. It is easy to see that the zero modes form the two-dimensional ECS algebra, as we show in appendix \ref{appendix:dimred}. Because these are the physical symmetries of the system, the quantum states must furnish a projective representation of the two-dimensional ECS—equivalently, a unitary irreducible representation of the QCS—and our formalism is accordingly adapted.

 The goal of this section is therefore to compute the quantum gravitational entanglement entropy of a spherical region in four-dimensional spherically symmetric spacetime, using the formalism of corner symmetries. For simplicity, we restrict our analysis to near-extremal Reissner–Nordström black holes, where the dilaton theory reduces exactly to JT gravity, leaving the generalization to arbitrary spherically symmetric spacetimes to future work. The causal diamond perspective of \cite{Banks:2021jwj,Gukov:2022oed} is also described throughout. 
 While the perspective offered by dimensional reduction is illuminating, the results can also be understood from the viewpoint of pure two-dimensional JT gravity.

\subsection{Dimensional reduction}\label{dimre}
We consider a general four-dimensional spacetime $M_4$ exhibiting spherical symmetry. In full generality, the corresponding metric can be expressed as
\begin{equation}\label{eq:sphericallysymmetricmetric}
    \dd s^2_{M_4} = g_{ab}(x^c)\, \dd x^a \dd x^b + \rho^2(x^c)\, \dd\Omega^2_S,
\end{equation}
where latin indices $a,b,c$ take values $0,1$, the function $\rho(x^c)$ is a scalar field independent of the angular coordinates, and $\dd \Omega^2_S$ represents the standard line element on the unit 2-sphere.
The Einstein–Hilbert action associated with this spherically symmetric spacetime is given by
\begin{equation}\label{eq:M4EHaction}
    S_{M_4} = \frac{1}{16\pi G_N} \int \dd^4 x\, \sqrt{-g_{M}}\, R_4,
\end{equation}
where $G_N$ denotes Newton's gravitational constant, $g_M$ is the determinant of the full spacetime metric \eqref{eq:sphericallysymmetricmetric}, and $R_4$ is the corresponding Ricci scalar.
The metric can be rewritten as a conformal transformation of the metric on a product space between a Lorentzian two-manifold $\tilde{M}_2$ and a sphere of radius $L$
\begin{equation}
    \dd s^2_{M_4} = \frac{\rho^2}{L^2}\qty(\tilde{g}_{ab}\dd x^a \dd x^b + L^2 \dd \Omega_S^2),
\end{equation}
where 
\begin{equation}\label{eq:gtildeab}
    \tilde{g}_{ab} = \frac{L^2}{\rho^2}g_{ab},
\end{equation}
and we have left the argument  $x^c$ implicit to improve readability.
The curvature of the metric in parenthesis then takes the simple form
\begin{equation}
    \tilde{R}_4 = \tilde{R}_2 + \frac{2}{L^2},
\end{equation}
where $\tilde{R}_2$ is the Ricci scalar associated with the two dimensional metric \eqref{eq:gtildeab}. Using the standard formula for the conformal transformation of Ricci scalars
\begin{equation}
     R_4 = L^2\qty(\rho^{-2}\tilde{R}_4 -6 \rho^{-3}\tilde{\Box}\rho),
\end{equation}
we can rewrite the action \eqref{eq:M4EHaction} in terms of the metric $\tilde{g}$,
\begin{Align}
    S_{M_4} &= \frac{1}{16\pi G_N}\frac{1}{L^2}\int_{M_4} \dd^4 x \sqrt{-\tilde{g}}\, \qty[\rho^2 \qty(\tilde{R}_2 + \frac{2}{L^2}) - 6 \rho \tilde{\Box} \rho]\\
    &= \frac{1}{4 G_N} \int_{\tilde{M}_2}\dd^2 x \sqrt{-\tilde{g}_2}\qty[\rho^2\qty(\tilde{R}_2+\frac{2}{L^2})+6 \qty(\tilde{\nabla}\rho)^2] - \frac{3}{4G_N}\int_{\partial \tilde{M}_2}\epsilon_{\partial\tilde{M}_2}n_a \tilde{\nabla}
    ^a \qty(\rho^2),
\end{Align}
where we have used the fact that $\tilde{R}_2$ and $\rho$ do not depend on angular variables to integrate over the sphere going from the first to the second line as well as integration by parts. Here, $\tilde{g}_2$ denotes the determinant of the metric defined in~\eqref{eq:gtildeab}, the boundary $\partial \tilde{M}_2$ is taken to be the black hole horizon with $\epsilon_{\partial \tilde{M}_2}$ denoting the volume form on it, and $n_a$ is the normal one-form to the boundary.
This action can be put into a more standard form of dilaton theories in two-dimensions
\begin{equation}\label{eq:twodimensionaldilatonaction}
    S_{M_4} = \int_{\tilde{M}_2}\dd^2 x \sqrt{-\tilde{g}}\qty[\Phi\qty(\tilde{R}_2 + \frac{2}{L^2}) + \frac32 \frac{(\tilde{\grad}\Phi)^2}{\Phi}] - 3 \int_{\partial \tilde{M}_2}\dd t\sqrt{-\tilde{g}_1}n_a \tilde{\nabla}^a \Phi,
\end{equation}
where $\Phi$ controls the area of the spherical region
\begin{equation}
    \Phi = \frac{\rho^2}{4 G_N}.
\end{equation}
Let us now comment on the boundary term. The boundary of the two dimensional geometry is exactly the spherically symmetric null hypersurface defined by the $\Phi = cste$ condition. Therefore, the gradient of $\Phi$ is normal to the boundary and the contraction with the null normal vanishes. The dimensional reduction of Einstein-Hilbert theory is therefore described by the bulk term in \eqref{eq:twodimensionaldilatonaction}.\par
In principle, one could proceed with the symplectic analysis of the action and compute the corner charges. We expect the resulting algebra to realize the two-dimensional ECS, and our treatment of entanglement entropy to remain applicable. However, for simplicity, we will focus here on the specific case of near-extremal Reissner–Nordström (NERN) black holes, which will simplify the action further. \par
The metric of NERN is
\begin{equation}\label{eq:RNmetricfull}
    \dd s^2 = -f(r)\dd t^2 +\frac{\dd r^2}{f(r)} + r^2 \dd \Omega^2_S,
\end{equation}
with 
\begin{equation}
    f(r) = \qty(1- \frac{2M}{r}+\frac{Q^2}{r^2}).
\end{equation}
The horizons for which the metric components are singular are at the locations
\begin{equation}
    r_\pm = M\pm \sqrt{M^2 -Q^2}.
\end{equation}
For near extremal black holes, $M = Q + \Delta M$ with $\abs{\Delta M} \ll \abs{Q}$, the horizon locations become
\begin{equation}
    r_\pm \approx Q\pm \sqrt{2Q\Delta M}.
\end{equation}
We are interested in the geometry near the outer horizon. Following \cite{Mertens:2022irh} (see also \cite{Hadar:2017ven}), we define a radial coordinate $\rho$ through
\begin{equation}
    r = Q + \rho, 
\end{equation}
such that, near the horizon, this coordinate scales as $\rho \sim \sqrt{\Delta M Q}$. The near-horizon geometry is then given by the metric
\begin{equation}
    \dd s^2 \approx -\frac{\qty(\rho^2-\rho_h^2)}{Q^2} \dd t^2 + \frac{Q^2}{\qty(\rho^2-\rho_h^2)} \dd \rho^2 + Q^2 \dd \Omega^2_S,
\end{equation}
with $\rho_h = \sqrt{2 \Delta M Q}$. This is the well known fact that the near-horizon geometry of near-extremal black holes is isomorphic to AdS$_2 \times S^2$, with $Q$ determining both the AdS length and the sphere radius. While the Schwarzschild coordinates are well suited to describe the exterior of the horizon, the horizon itself is better described in global coordinates
\begin{align}
    x &= \pm Q\cosh(\frac{\rho_h t}{Q^2})\sqrt{\frac{\rho^2}{\rho_h^2}-1},\label{eq:xfotrho}\\
    \tan(\frac{\tau}{Q}) &= \frac{\rho_h}{\rho} \sinh(\frac{\rho_h t }{Q^2})\sqrt{\frac{\rho^2}{\rho_h^2}-1},\label{eq:taufotrho}
\end{align}
where the positive square root in the first equation corresponds to the right patch of AdS$_2$, which corresponds to the region described by the Schwarzschild coordinates, and the negative root is the extension to the two-sided AdS$_2$ black hole. From the second equation we see that the $\tau$ coordinates is restrained to $\tau\in \qty[-Q \frac{\pi}{2},Q\frac{\pi}{2}]$. In those coordinates, the metric becomes
\begin{equation}\label{eq:ads2metricglobalcoordinate}
    \dd s^2 = -\qty(1 + \frac{x^2}{Q^2})\dd \tau^2 + \frac{1}{\qty(1+\frac{x^2}{Q^2})}\dd x^2 + Q^2 \dd \Omega^2_S.
\end{equation}
From equation \eqref{eq:xfotrho} and \eqref{eq:taufotrho}, we see that the black hole horizon gets mapped to the bifurcate horizon $x=0,\tau=0$ and the metric coefficients in \eqref{eq:ads2metricglobalcoordinate} are well-defined at that point.\par
In order to realize the isomorphism dynamically, we perform the dimensional reduction of the NERN. The metric \eqref{eq:RNmetricfull} is of the form \eqref{eq:sphericallysymmetricmetric} with the identification $\Phi(t,r) = \frac{r^2}{4 G_N}$. We can put it in the conformally rescaled form using the length scale $L = Q$. The two-dimensional metric \eqref{eq:gtildeab} is given by
\begin{equation}
   \qty(\tilde{g}_{ab}) = \frac{Q^2}{r^2} \qty(-f(r) \dd t^2 + \frac{1}{f(r)}\dd r^2).
\end{equation}
Using the above, we can compute the kinetic term in the action \eqref{eq:twodimensionaldilatonaction}
\begin{equation}
    \frac{\qty(\tilde{\nabla}\Phi)^2}{\Phi} = \frac{Q^2-2 M r + r^2}{G Q^2}.
\end{equation}
In the near-horizon limit, the leading term of the above is
\begin{equation}
    \frac{\qty(\tilde{\nabla}\Phi)^2}{\Phi} \approx - \sqrt{2 Q \Delta M} \frac{\Delta M}{G Q^2},
\end{equation}
where we denoted $M = Q + \Delta M$. The kinetic term is thus subdominant in the near-extremal limit and can be dropped out of the action \eqref{eq:twodimensionaldilatonaction}. We are then left with the simple JT gravity action \cite{Grumiller:2002nm}
\begin{equation}\label{eq:JTaction}
    S_{RN} = \int_{\tilde{M}_2} \dd^2 x \sqrt{-\tilde{g}} \Phi\qty(\tilde{R}_2 + \frac{2}{Q^2}).
\end{equation}
Before proceeding, we also note that a similar correspondence between the near-horizon geometry of a causal diamond in flat Minkowski space and the JT gravity action was established in~\cite{Gukov:2022oed}. There, the value of the dilaton on the boundary of the causal diamond is associated with the area of that boundary
\begin{equation}\label{eq:causaldiamondbc}
    \Phi\eval_{\partial \tilde{M}_2} = \frac{L^2}{4 G_N},
\end{equation}
where $L$ denotes the size of the diamond. Our formalism is thus naturally suited also to the description of entanglement entropy associated with causal diamonds in flat spacetime.
\subsection{Corner charges and entanglement entropy}\label{cocha}
The starting point of the symplectic analysis is the action given in equation \eqref{eq:JTaction}, which we rewrite here with simplified notation for clarity 
\begin{equation}
      S_{\mathrm{JT}} = \int_{M}\dd^2 x \sqrt{-g}\, \Phi\qty(R + \frac{2}{Q^2})=\int_M \mathcal{L}\dd^2 x.
\end{equation}
The variation of the Lagrangian density gives
\begin{equation}
    \delta \mathcal{L} = E_g^{ab}\delta g_{ab} + E_\Phi \delta \Phi + \partial_a \theta^a, 
\end{equation}
with
\begin{align}
    E^{ab}_g &= \sqrt{-g}\qty(\nabla^a \nabla^b \Phi -\nabla^2 \Phi g^{ab} + \frac{\Phi}{Q^2} g^{ab}),\label{eq:eomJTg}\\
    E_\Phi &= \sqrt{-g}\qty(R + \frac{2}{Q^2}),\label{eq:eomJTphi}\\
    \theta^a &=  \sqrt{-g}\qty(\Phi g^{ab}\nabla^c - \Phi g^{bc}\nabla^a + \nabla^a \Phi g^{cb}-\nabla^c \Phi g^{ab})\delta g_{bc}.\label{eq:symplecticpotJT}
\end{align}
The solutions to the equations of motion correspond to two-sided AdS$_2$ black holes. In global coordinates, they can be written
\begin{align}
    \dd s^2 &= -\left(1 + \frac{x^2}{Q^2}\right) \dd \tau^2 + \left(1 + \frac{x^2}{Q^2}\right)^{-1} \dd x^2, \label{eq:metricsolution}\\
    \Phi(\tau,x) &= \Phi_h \sqrt{1 + \frac{x^2}{Q^2}} \cos\left(\frac{\tau}{Q}\right), \label{eq:dilatonsolution}
\end{align}
where $\Phi_h$ denotes the value of the dilaton on the black hole horizon which is defined by the $x=\pm M \tan(\frac{\tau}{M})$ curves. In particular, we know from  our previous discussion that the horizon of the original NERN black hole sits at the point $x=\tau=0$. This allows us to identify
\begin{equation}\label{eq:identphihM}
    \Phi_h = \frac{Q^2}{4 G_N} + \frac{Q^{\frac32}}{G_N}\sqrt{\frac{\Delta M}{2}} + \mathcal{O}\qty(\Delta M).
\end{equation}
Similarly, in the causal diamond perspective, the identification follows from equation \eqref{eq:causaldiamondbc}, 
\begin{equation}\label{eq:causaldiamondphih}
    \Phi_h = \frac{L^2}{4 G_N}.
\end{equation}
\par
We now turn to the computation of the corner charges. In order to define our symplectic structure, we need to choose a codimension-1 hypersurface. Since we want the corner to correspond to the black hole horizon, we choose the $t=0$ Cauchy slice in the original four-dimensional spacetime, which corresponds to the $\tau = 0$ line in the AdS$_2$ geometry, see \cite{Harlow:2018tqv} and \cite{Grumiller:2021cwg} for more details.
From the perspective of pure two-dimensional JT gravity, the glued segment corresponds to the $\tau = 0,x\in \qty(-\infty,\infty)$ line and the corner is the entangling surface between the left and right universes located at $p_S = (0,0)$ in global coordinates.
Within the extended phase space formalism described in Section~\ref{sec:quantumcorners}, the corner charge is given by the pullback to the corner of the quantity $\mathrm{Q}_\xi$, defined in equation~\eqref{eq:noethercurrent}. Since the Lagrangian vanishes on-shell, the only contributing term is the contraction of the symplectic potential \eqref{eq:symplecticpotJT} with the field space vector field \eqref{eq:fieldspacevectorfield}. We find
\begin{equation}\label{JTnoethercharge}
Q_\xi = -2\frac{\epsilon^{ab}}{\sqrt{-g}}\left(\Phi\qty(\, \partial_a g_{bc}\xi^c + g_{bc}\partial_a\xi^c) -2 \partial_a\Phi \,g_{bc}\xi^c \right)\eval_{p_S}. 
\end{equation}

In order to identify the algebra and its generators, we follow \cite{Ciambelli:2021vnn} and expand the fields and diffeomorphisms around the corner.
 We find
\begin{equation}\label{JTnoetherchargecoordinates}
    Q_\xi = \xi_{(0)}^c t_c + \xi^c_{(1)a} N^a_c,
\end{equation}
where 
\begin{Align}
    t_c &= -2\frac{\epsilon^{ab}}{\sqrt{-g^{(0)}}}\left(\Phi^{(0)}g_{bca}^{(1)}-2\Phi_a^{(1)}g_{bc}^{(0)}\right),\\
    N^a_c &= -2\frac{\epsilon^{ab}}{\sqrt{-g^{(0)}}}\Phi^{(0)}g_{bc}^{(0)},\label{eq:classicalsl2rcharge}
\end{Align}

where the $(0)$ and $(1)$ subscript respectively denote the fields and their first derivatives evaluated at the corner. Since this holds for any diffeomorphism, the components $\xi_{(0)}$ parametrize the Lie algebra of translations $\mathbb{R}^2$, while the components $\xi_{(1)}$ parametrize the general linear group $\mathrm{GL}(2,\mathbb{R})$.
The covariant phase space induces the following non-vanishing Poisson bracket on the field space
\begin{equation}
    \qty{N^a_b,N^c_d} = \delta^c_b N^a_d - \delta^a_d N^c_b,\qquad \qty{N^a_b,t_c} = -\delta^a_c t_b +\frac12 \delta^a_b t_c.    
\end{equation}
Since the matrix $N^a_{c}$ is traceless, it generates only the $\mathfrak{sl}(2,\RR)$ subalgebra of the general linear algebra. This confirms that the corner symmetries of JT gravity form the ECS group. Since the translations will acquire a central charge in the corresponding quantum theory, we can associate the charges to the Hermitian operators of the representation \eqref{Hermitianbasis} in the following way
\begin{Align}\label{eq:chargeoperatorcorrespondence}
    N^0_0 &\rightarrow D, \quad N^1_0\rightarrow -K\quad N^0_1\rightarrow H\\
    &t_0 \rightarrow X,\qquad t_1\rightarrow P.
\end{Align}
More rigorously, this is the result of the momentum map, as we explain in appendix \ref{appendix:momentummaps}.

We now want to connect the value of the classical charges with the quantum representations. In order to do so, we need to express the Casimir operator in terms of the classical variables appearing in the charges. The Casimir operator \eqref{eq:qcscubiccasimir} is an inherently quantum operator whose definition relies on the fact that the translations are not commuting. To relate it to the classical analysis, we can use that the expectation value of quantum operators in a coherent state is given by the value of the associated classical observable. If we reinstate units by restoring the explicit factor of $\hbar$, the expectation value of a product of operators equals the product of the corresponding classical values, up to terms of order $\hbar$ which represent the quantum correction. In other words, this is the standard procedure of writing correlators as tree-level (classical) values plus loop corrections.
In the classical limit $\hbar \mapsto 0$, the expectation value of the cubic Casimir \eqref{eq:qcscubiccasimir} in the coherent state \eqref{cost} with the operator identification as in \eqref{eq:chargeoperatorcorrespondence} can therefore be written as
\begin{equation}
\expval{\mathcal{C}_{\mathrm{QCS}}}_{\zeta,\gamma_\zeta} = 4 c \Phi_h^2 = \frac{c Q^4}{4 G_N^2\hbar^2} + \frac{c Q^\frac72 \sqrt{2} \sqrt{\Delta M}}{G_N^2\hbar^2} + \mathcal{O}\qty(\Delta M) + \mathcal{O}\qty(\hbar) ,
\end{equation}
where we used the classical values of the charges \eqref{eq:classicalsl2rcharge}, the particular solutions \eqref{eq:metricsolution} and \eqref{eq:dilatonsolution}, and the identification \eqref{eq:identphihM} where the factor of $\hbar$ was reinstated.

On the other hand, the expectation value of the Casimir operator in any normalized state is given by \eqref{eq:casimirqcsrep}. In the classical limit where the length scales of the problem are much larger than the Planck scale $Q \gg \Delta M\gg \sqrt{G_N\hbar}$, this allows us to identify the representation parameter as
\begin{equation}
    s = \frac{Q^2}{2 G_N\hbar}\qty(1 + 2\sqrt{2}\sqrt{\frac{ \Delta M}{Q}}),
\end{equation}
where we dropped higher terms in $\Delta M$ and in $Q^2/(G_N\hbar )$. We see that this classical limit thus naturally corresponds to the large $s$ representations. 

Now consider the classical state $\ket{\psi_{cl}\qty(\pi s)}$ defined in \eqref{eq:classicalstate}. The entanglement entropy of that state has leading order
\begin{equation}
    S^{\psi_{cl}} \approx \frac{\pi Q^2}{G_N\hbar }\qty(1 + 2\sqrt{\frac{ 2\Delta M}{Q}}).
\end{equation}
This exactly reproduces the NERN area law
\begin{equation}
    S^{\psi_{cl}} = \frac{4\pi r_+^2}{4 G_N\hbar}+\mathcal{O}\qty(\Delta M^\frac32),
\end{equation} 
thereby justifying its definition as a classical state. Additionally, the same state in the Minkowski causal diamond case \eqref{eq:causaldiamondphih} also has the expected area law in the classical limit
\begin{equation}
  S^{\psi_{cl}} \approx \frac{4 \pi L^2}{4 G_N\hbar }.
\end{equation}
We also note that from the asymptotic behaviour \eqref{eq:asymptoticentropyclassicalstate}, the next to leading order has the expected logarithmic form of quantum corrections to the area law. 

Therefore, we have shown how applying our general formalism of $2$d corners to JT gravity precisely reproduces the expected area law in the entanglement entropy of coherent states of the QCS. We take this as a confirmation of the corner proposal, and of its capacity of linking quantum effects with semiclassical limits faithfully. This is incidentally also a motivation to generalize the study of representations and states to higher dimensions, without restricting to spherically-symmetric configurations. 

\section{General Quantum Gravity States}\label{genqu}

The most general quantum state associated with a corner is a superposition of states belonging to different representations. To construct such states, it is necessary to suitably sum over all representations. A natural mathematical framework for this summation is provided by the Plancherel measure, which arises in the direct integral decomposition of square-integrable functions on the $\mathrm{QCS}$ group with respect to the Haar measure
\begin{equation}
    L^2(\mathrm{QCS}) \cong\int_{\widehat{\mathrm{QCS}}} \dd\mu(s,c)\, \mathrm{HS}(\mathcal{H}^{(s,c)}) \cong \,\int_{\widehat{\mathrm{QCS}}} \dd\mu(s,c) \qty(\mathcal{H}^{(s,c)}\otimes \bar{\mathcal{H}}^{(s,c)}),
\end{equation}
where $\widehat{\mathrm{QCS}}$ denotes the unitary dual of the group, $\dd\mu(s,c)$ is the Plancherel measure, and $\mathrm{HS}(\mathcal{H}^{(s,c)})$ denotes the Hilbert-Schmidt space over the representation space $\mathcal{H}^{(s,c)}$.
Note that this formula can be seen as the non-compact generalization of the Peter--Weyl theorem. 
A general state can then be written as
\begin{Align}\label{eq:generalstate}
    \ket{\Phi} &= \int \dd \mu(s,c)\, \Phi(s,c) \intpE{p}{E} \psi(E)\phi(p)\ket{s;E}\ket{c;p},
\end{Align}
where we now explicitly indicate the representation parameters $s$ and $c$ associated with the $\spl{2}$ and the Heisenberg states, respectively. The Plancherel measure, together with the associated Fourier inversion theorem, provides an orthogonal decomposition; in particular, operators belonging to distinct sectors are orthogonal. The completeness relation of the $\ket{s, c; E, p}$ states in a given Hilbert space now acts as a projector onto that Hilbert space
\begin{equation}
    \intpE{p}{E} \ket{s,c;E,p}\bra{s,c;E,p} = P_{(s,c)},
\end{equation}
and the completeness relation on the total Hilbert space reads
\begin{equation}
    \int \dd \mu(s,c)\, P_{(s,c)} = \mathds{1},
\end{equation}
which is precisely the decomposition of a Hilbert space into distinct orthogonal spaces. The above completeness relation implies the normalization\footnote{If the representation is labeled by a discrete parameter, the delta function becomes a Kronecker delta.}
\begin{equation}
\braket{s',c';E',p'}{s,c;E,p} = \delta_\mu(s-s')\delta_\mu(c-c') \delta(E-E')\delta(p-p'),
\end{equation}
where $\delta_\mu$ is the adapted Dirac delta such that
\begin{equation}
    \int \dd \mu(s,c) \,\Phi(s,c) \delta_\mu(s-s') \delta_\mu(c-c') = \Phi(s',c'),
\end{equation}
for any function $\Phi(s,c)$. The general state is then normalized if the wave function is normalized with respect to the Plancherel measure
\begin{equation}
    \braket{\Phi} = \int \dd \mu(s,c) \abs{\Phi(s,c)}^2 = 1. 
\end{equation}
\par
We can now construct a general corner state simply as
\begin{equation}
    \ket{\Phi} = \int \dd \mu(s,c) \,\Phi(s,c)\intpE{p}{E}  \psi(E)\phi(p) \ket{s;E}\ket{c;p}.
\end{equation}
Next, we can split it into subsystems using the doubling map
\begin{equation}
    \ket{\Phi}\longmapsto \int \dd \mu(s,c)\, \Phi(s,c)\intpE{p}{E} \psi(E)\phi(p) \ket{s;E}_L\ket{c;p}_L\otimes \ket{s;E}_R\ket{c;p}_R.
\end{equation}
Note that the presence of the Casimirs in the maximally commuting subalgebra now becomes essential. It is reflected in the fact that the tensor product state has the same value of $s$ and $c$ on the left and on the right. Considering the density operator associated with the doubled state and tracing over the right degrees of freedom yields 
\begin{equation}
    \rho_L = \int \dd \mu(s,c)\, \dd E \abs{\Phi(s)}^2 \abs{\psi(E)}^2 \abs{\phi(p)}^2  \ket{s;E}\ket{c;p}\bra{s;E}\bra{c;p}.
\end{equation}
The associated entanglement entropy is given by
\begin{equation}\label{eq:entanglemententropygeneralstate}
    S^{\psi,\phi} =\int \dd \mu(s,c) \,  \abs{\Phi(s,c)}^2 \left(-\ln(\abs{\Phi(s,c)}^2) + S^{\psi,\phi}(s,c)\right),
\end{equation}
where $S^{\psi,\phi}(s,c)$ denotes the entanglement entropy of the representation $(s,c)$ associated with the specific state determined by the wave function $\psi(E)\phi(p)$, as computed in the previous sections. This quantity determines the contribution from each representation. 

In particular, using the Boltzmann interpretation of entropy, we can view it as a measure of the effective dimension of the representation in a particular state
\begin{equation}
    S^{\psi,\phi}(s,c) = \ln( d_{\mathrm{eff}}(s,c)).
\end{equation}
Plugging this back into the total entropy yields
\begin{equation}
    S^{\psi,\psi} = \int \dd \mu(s,c)\, \abs{\Phi(s,c)}^2\qty(-\ln(\abs{\Phi(s,c)}^2)+ \ln(d_{\mathrm{eff}}(s,c))).
\end{equation}
This result bares a striking resemblance to the entanglement entropy of non-Abelian gauge theories of Donnelly \cite{Donnelly:2014gva}
\begin{equation}\label{eq:Donnellyformula}
    S = \sum_R \abs{\psi(R)}^2\left(-\ln(\abs{\psi(R)}^2 + 2\ln(\mathrm{dim}R))\right),
\end{equation}
where, in our case, the role of the representation dimension of the compact gauge group is played by the exponential of the entanglement entropy of each representation
\begin{equation}
    \mathrm{dim}R \sim \exp(S^{\psi,\phi}(s,c)).
\end{equation}
Equation \eqref{eq:entanglemententropygeneralstate} is therefore a direct generalization of Donnelly's formula for the non-compact case. The fact that the entropy only depends on the modulus $\abs{\Phi(s,c)}$ and not on the relative phase is a consequence of the superselection rule, according to which states in different representation cannot be in superposition. The second term in \eqref{eq:Donnellyformula} corresponds to the contribution from boundary degrees of freedom, which introduce a maximally mixed state of dimension $\mathrm{dim}R$ at each corner. The factor of 2 accounts for the two endpoints of the segment. This interpretation aligns with the understanding that the entanglement entropy $S^{\psi,\phi}(s,c)$ originates from the presence of non-trivial representations of the corner symmetries and is directly related to the gluing procedure, which produces mixed states. In the present work, we consider only a single corner, and the factor of 2 is therefore absent.

We now focus our attention on a specific example that has illuminating physical consequences.
Since a complete description of the Plancherel measure for the QCS is still under investigation, we begin by restricting  to the special linear component, which corresponds to boundary-preserving diffeomorphisms. Although certain aspects of the QCS structure are absent in this setting, we expect the general computations to exhibit similar features. Finally, before proceeding, we note that only the positive discrete series are considered in the present analysis.

The Plancherel measure for the discrete series of the universal cover of $\spl{2}$ is given by
\cite{PUKANSZKY1964}
\begin{equation}
    \dd \mu(s) = \frac{2}{\pi^2}s\,\dd s,
\end{equation}
with $s \in \RR$. We consider a Gaussian wave function on the representation space, given by
\begin{equation}\label{eq:generalstategaussianwavefunction}
   \Phi(s) = \mathcal{N} \, e^{-\frac{(s - s_0)^2}{2}},
\end{equation}
and focus on the limit of large $s_0$, where the wave function is sharply peaked at large values of the representation parameter $s$, and thus contributes significantly only in that regime. The normalization constant can be calculated using the Plancherel measure
\begin{equation}
    \mathcal{N} = \sqrt{\frac{\pi^{\frac32}}{2 s_0}} + \mathcal{O}\qty(s_0^{-\frac32}).
\end{equation}
For the state in each representation, we will consider the classical coherent state \eqref{eq:classicalstate}. Since the integral only contributes at large values of $s$, we can consider the asymptotic expansion \eqref{eq:asymptoticentropyclassicalstate}.
We then split equation \eqref{eq:entanglemententropygeneralstate} in two pieces
\beq
S^{\psi,\phi}&=&S_{\text{bulk}}^{\psi,\phi}+S_{\text{boundary}}^{\psi,\phi}\\
S_{\text{bulk}}^{\psi,\phi}&=&-\int \dd \mu(s,c) \,  \abs{\Phi(s,c)}^2 \ln(\abs{\Phi(s,c)}^2)\\
S_{\text{boundary}}^{\psi,\phi}&=&\int \dd \mu(s,c) \,  \abs{\Phi(s,c)}^2 S^{\psi,\phi}(s,c)
\eeq
and compute their leading order terms
\begin{Align}\label{eq:firstandsecondterm}
    S_{\text{bulk}}^{\psi,\phi}= \ln(s_0) + \mathcal{O}(1),\qquad
     S_{\text{boundary}}^{\psi,\phi}=s_0 + \mathcal{O}\qty(\ln(s_0)).
\end{Align}
In the specific case in which the parameter $s$ corresponds to the area of the four-dimensional Reissner-Nordström black hole, we can extract the following physical interpretation from the result above. The general state defined by the wave function \eqref{eq:generalstategaussianwavefunction} can be seen as a superposition of fuzzy area states sharply peaked around the classical value $s_0$. Looking at equation \eqref{eq:firstandsecondterm}, we observe that the area law emerges from the boundary contribution. This is an expected universal result which has also been realized in loop quantum gravity, string theory and matrix models \cite{Donnelly:2008vx,Delcamp:2016eya,Bianchi:2024aim,Donnelly:2020teo,Frenkel:2023aft}. 

In conclusion, we have seen how the general states constructed from the corner proposal have an associated entanglement entropy which can faithfully reproduce the expected area-law behavior. This result helps in promoting the quantum aspects of the proposal to more solid grounds, and in demonstrating how this approach to quantum gravity yields promising insights and far-reaching consequences.

\section{Conclusions}\label{concl}

In this manuscript, we explored the corner proposal further by studying the quantum information properties of the corner states. The corner proposal is based on the understanding of the representations of the corner algebra
\beq
\mathrm{Diff}\qty(S)\ltimes \qty(\spl{2}\ltimes \RR^2)^S.
\eeq
The study of the representations of this group can be a daunting task. We thus initiated in \cite{Ciambelli:2024qgi} a simpler pathway, focused on the study of its $2$-dimensional realization
\beq
\spl{2}\ltimes \RR^2.
\eeq
At the quantum level, one must include all possible central extensions of this group, leading to the quantum corner symmetry group
\beq
\reallywidetilde{\mathrm{SL}}\qty(2,\RR)\ltimes H_3.
\eeq

We reviewed how these groups come about in section \ref{coalg}, and the representations of the QCS in section \ref{reglu}, which is based on \cite{Varrin:2024sxe} (see also \cite{Ciambelli:2022cfr, Neri:2025fsh}). In section \ref{reglu}, we also discussed how one can break the spacetime into two pieces by splitting the corner group --a procedure we called the doubling map--, and conversely how one can glue together spacetime subregions in this algebraic setting.

Representations of the QCS are associated with gravitational configurations at corners. Moreover, given our doubling map, we can algebraically restrict our attention to a subregion by consistently tracing over the other degrees of freedom. This allows us to straightforwardly make contact with quantum information concepts such as the von Neumann and Rényi entanglement entropy. This prompted us to study in section \ref{entent} specific states in these representations with relevant physical content. We started with the vacuum state in section \ref{vasta}, which we interpreted as the absence of classical geometry, and computed its entanglement entropy once reduced to a subregion, finding a quantum geometric version of the Unruh effect. Then, with the intention of having states realized in the semiclassical limit, we focused on coherent states and their entropy in section \ref{costa}. QCS coherent states were not known, so we constructed them using the semi-direct nature of the group and its representations, and the already-known coherent states of the two subgroup forming the QCS. Eventually, we analyzed the Rényi entropy of these states in section \ref{sec:rényientropy}. Interestingly, we found that vacuum and coherent states have the same modular fluctuations which, for certain values of the representation parameter, is controlled by the expectation value of the modular Hamiltonian itself. 

One of the main shortcomings of the corner proposal so far was its lack in relating to semiclassical limits and in having explicit realizations in simple quantum gravitational systems. We filled this gap in section \ref{sec:arealaw}, in which we applied our technical formalism to $2$-dimensional JT gravity. After reviewing in section \ref{dimre} how the latter is the result of dimensionally-reducing the near horizon dynamics of near extremal spherically-symmetric black holes in $4$d, we applied the corner proposal to it in section \ref{cocha}. We thus constructed the corner charges and mapped them to the coalgebra of the QCS. Then, we demonstrated that the entanglement entropy of the classical coherent states constructed from purely algebraic considerations in the previous section exactly reproduces the area law behavior in the semiclassical limit. This was one of the main results of this manuscript, linking the abstract corner proposal formulation of $2$d gravity with concrete semiclassical predictions, and reproducing the area-law scaling without resorting to a background structure. Indeed, our coherent states are quantum geometric configurations of corners. As such, they do not rely on a classical background geometry, they are geometry themselves. Eventually, we considered in section \ref{genqu} more general quantum states constructed as superposition of states in different QCS representations. There, we found a formula for the entanglement entropy precisely generalizing the result obtained by Donnelly \cite{Donnelly:2014gva} in non-Abelian gauge theories. 

Our results open the door to many  future directions, mostly separated into two macro area.
\begin{itemize}
    \item The first one is the continuation of this $2$-dimensional analysis. Here, it would be interesting to characterize more general quantum states and their appearance -- or not -- in the semiclassical description of the theory. A question we only partially addressed was the saturation of the modular Hamiltonian fluctuations, in the spirit of \cite{Verlinde:2019xfb, Verlinde:2019ade, Verlinde:2022hhs}. This deserves further attention, as the link between our work and the understanding of the causal diamond fluctuations could certainly being rewarding. Another question we raised concerns the geometric quantization of JT gravity. We wonder how to relate our algebraic approach to the one of \cite{Maldacena:2016upp}, in which the boundary reparametrization mode and Schwarzian theory (see \cite{Mertens:2022irh}) are utilized. We expect that the cubic QCS Casimir will play an important role here, yet to be unveiled.
    \item The second macro area is the study of the corner proposal in higher dimension. Indeed, although our studies so far bore interesting fruits, the addition of the internal diffeomorphisms of the corner is a hard yet important task to face. While this challenge has already been initiated in \cite{Donnelly:2020xgu, Ciambelli:2022cfr, Donnelly:2022kfs}, these studies confined their attention to the ECS group itself (or subgroups), without studying the possible central extensions. The latter are fundamental in the quantum representation theory, and so we plan to study them in the future. By retracing the steps taken in $2$d, our strategy, after studying the central extensions, is to construct the doubling and gluing procedures, and then to  construct quantum states. Then, we can scrutinize the quantum informational features of these states. Eventually, a concrete example will certainly be beneficial, to test our algebraic and abstract construction.
\end{itemize}
To summarize, this paper plays a valuable role in solidifying the corner proposal and establishing its connection to concrete quantum gravity predictions, yet much remains to be uncovered in this rich and promising direction of research.

\paragraph{Acknowledgments} We are thankful to Ivan Agullo, Andreas Blommaert, Laurent Freidel, Florian Girelli, Temple He, Josh Kirklin,  Rob Leigh, Etera Livine, Juan Maldacena, Djordje Minic, Rob Myers, Giulio Neri, Simone Speziale, and Kathryn Zurek for discussions and related projects.
Research at Perimeter Institute is supported in part by the Government of Canada through the Department of Innovation, Science and Economic Development Canada and by the Province of Ontario through the Ministry of Colleges and Universities. This work was supported by the Simons Collaboration on Celestial Holography. 

\appendix
\section{Dimensional reduction of corner symmetries}\label{appendix:dimred}
Consider Einstein-Hilbert theory on a four-dimensional spacetime manifold $M$. The corner $S$ is defined as an embedding of a two-dimensional surface into the manifold \cite{Ciambelli:2021vnn}
\begin{equation}
    \phi: S \longrightarrow M.
\end{equation}
The normal space to the corner is the vertical sub-bundle
\begin{equation}
    \mathsf{V} = \mathrm{ker}\qty(\dd \phi).
\end{equation}
In order to discuss the splitting of the spacetime tangent bundle into vertical and horizontal part, we introduce adapted coordinates on the manifold
\begin{equation}
    y^\mu = \qty(u^a,x^i),
\end{equation}
where $\mu = 0,1,2,3$, while $u^a, a=0,1$ are the normal coordinates to the corner and $x^i, i=2,3$ are the tangential coordinates. In terms of those, the vertical bundle
is simply
\begin{equation}
    \mathsf{V} = \mathrm{span}\qty{\partial_a}.
\end{equation}
One can now introduce the normal one forms
\begin{equation}
    n^a = \dd u^a - \omega^a_i(u,x) \dd x^i, 
\end{equation}
where $\omega^a_i$ plays the role of an Erhesmann connection which represents the ambiguity in the decomposition of the tangent bundle
\begin{equation}
    \mathrm{T}M = \mathsf{V} \oplus \mathsf{H},
\end{equation}
where $\mathsf{H}$ is defined by
\begin{equation}
    \mathsf{H} = \mathrm{ker}\qty(n^a) = \qty{\xi \in \mathrm{T}M \mid n^a(\xi) = 0}.
\end{equation}

We now introduce the metric in adapted coordinates
\begin{equation}\label{eq:metricadaptedcoordinates}
    \dd s^2 = h_{ab}(u,x) n^a n^b + \gamma_{ij}(u,x)\dd x^i \dd x^j.
\end{equation}
Note that for Einstein--Hilbert theory, the corner charges at a finite-distance corner associated with a diffeomorphism \(\xi \in \mathrm{T}M\) are given by \cite{Ciambelli:2021vnn}
\begin{equation}\label{eq:EHnoethercharge}
    Q_{\xi} = \int_S \mathrm{vol}_S \qty(\xi^a_{(1)b} N^b_a + \xi^i_{(0)} b_i + \xi^a_{(0)} p_a),
\end{equation}
 where $\mathrm{vol}_S$ is the volume form on the corner and
\begin{align}
    N^b_a &= \sqrt{-\mathrm{det}h^{(0)}}h^{bc}_{(0)}\epsilon_{ca},\label{sl2rcharges}\\
    b_i &= -N^b_a \omega^a_{(1)ib},\label{cornerdiffcharges}\\
    p_a &= \frac12 N^a_c h^{cb}_{(0)}\qty(h^{(1)}_{dba}-h^{(1)}_{dab}),\label{normaltranslationcharges}
\end{align}
where the subscripts \((0)\) and \((1)\) respectively indicate the evaluation of the fields or their first derivative at the corner. It is straightforward to see that the \( b_i \) charge is associated with infinitesimal diffeomorphisms on the corner, the traceless \( N^a_b \) corresponds to the \(\mathfrak{sl}(2)\) algebra, and \( p_a \) to normal translations. The corner symmetry group of four-dimensional Einstein--Hilbert theory is therefore indeed the ECS group \eqref{eq:ecsgroup}.\par
We now consider a spherically symmetric spacetime and restrict to diffeomorphisms that preserve this symmetry. In that case, we have a global direct product structure
\begin{equation}
    M = M_2 \times S^2,
\end{equation}
or, in other words, there are no mixing between the normal and horizontal coordinates. In particular comparing the metrics \eqref{eq:sphericallysymmetricmetric} and \eqref{eq:metricadaptedcoordinates}, we see that the spherically symmetric metric can be written globally in an adapted coordinate system such that $\omega^a_i = 0$. This means that the Ehresmann connection is trivial and the corner diffeomorphism charges \eqref{cornerdiffcharges} vanish. Furthermore, the metric coefficients appearing in the \(\mathfrak{sl}(2)\) and \(\mathbb{R}^2\) charges do not depend on the angular coordinates. Since we have also restricted to diffeomorphisms that are independent of the angular coordinates, we can write the Noether charge \eqref{eq:EHnoethercharge} as
\begin{equation}
    Q_\xi = A_S \qty(\xi^a_{(1)b} N^b_a + \xi^a_{(0)} p_a),
\end{equation}
where $A_S$ is the area of the corner which can be absorbed by a field redefinition. The above then corresponds exactly to the Noether charges of two-dimensional Einstein-Hilbert gravity that generates the two-dimensional symmetry group \eqref{eq:twodimensionalecs}. 

\section{Momentum map}\label{appendix:momentummaps}
In order to connect the representation theory with the dynamical charges of JT gravity \eqref{JTnoethercharge}, we introduce the concept of momentum map. The momentum map links functionals of the classical field space $\Gamma$ to the coalgebra
\begin{equation}
    \mu : C^{\infty}\qty(\Gamma) \longrightarrow \mathfrak{ecs}^*,
\end{equation}
such that 
\begin{equation}
    \pairing{\mu(\varphi)}{X} = Q_X, \quad \forall X\in \mathfrak{ecs},
\end{equation}
where the bracket denotes the pairing between the algebra and its dual, $Q_X$ is the charge associated with the Hamiltonian vector fields generated by $X$ \eqref{JTnoethercharge}, and where we denoted $\varphi$ a general element of the field space configuration.
We denote the basis of the $\mathfrak{ecs}$ algebra with $\qty{J\updown{\mu}{\nu}, P_\mu},\, \mu,\nu=1,2$,
where $J\updown{\mu}{\nu}$ are the generators of $\spl{2}$ and $P_\mu$ generates the normal translations. Introducing the dual basis $\qty{\tilde{J}\downup{\mu}{\nu}, \tilde{P}^\mu}$ such that
\begin{equation}
    \pairing{\tilde{J}\downup{\mu}{\nu}}{J\updown{\rho}{\sigma}} = \delta^\rho_\mu \delta^\nu_\sigma, \quad \pairing{\tilde{P}^\mu}{P_\nu} = \delta^\mu_\nu,
\end{equation}
we can write any element of the coalgebra as
\begin{equation}
   q = \tilde{j}\downup{\mu}{\nu}\tilde{J}\downup{\nu}{\mu} + \tilde{p}_\mu \tilde{P}^\mu.
\end{equation}
The pairing between the $\mathrm{ecs}$ element generated by a charged diffeomorphism
\begin{equation}
    X_\xi = \xi^{\rho}_{(1)\nu} J\updown{\nu}{\rho} + \xi^{\mu}_{(0)}P_\mu,
\end{equation}
and the coalgebra element is simply expressed as
\begin{equation}
    \pairing{q}{X_\xi} = \tilde{j}\downup{\rho}{\nu}\xi^{\rho}_{(1)\nu} + \tilde{p}_\mu \,\xi^{\mu}_{(0)}.
\end{equation}
Using the defining equation of the momentum map and the Noether charge \eqref{JTnoetherchargecoordinates}, it follows that
\begin{equation}\label{momentummap}
    \mu(N^\mu_\rho) = \tilde{j}\downup{\rho}{\mu},\quad \mu(t_\mu) = \tilde{p}_\mu.
\end{equation}
We have thus successfully established a connection between the charges of JT gravity and the coalgebra structure of the extended corner symmetry group. The remaining challenge is to elucidate the precise link between this structure and the corresponding representation theory. 

The expressions in~\eqref{momentummap} associate coordinates on the coalgebra with the charge components $N^\mu_\rho$ and $t_\mu$. The momentum maps therefore play the role of coordinate functions corresponding to the dual basis of the Lie algebra, defined as
\begin{Align}
\mathcal{J}_X: \mathfrak{ecs}^* &\longrightarrow \RR,\\
    q &\mapsto \mathcal{J}_X(q) = \langle q, X \rangle.
\end{Align}
We can thus make the following identifications
\begin{equation}
    \mu(N^{\mu}_\rho) \rightarrow \mathcal{J}_{J\updown{\mu}{\rho}}, \quad \mu(t_\mu) \rightarrow \mathcal{J}_{P^\mu}.
\end{equation}
Upon quantization of the coadjoint orbits, these coordinate maps are promoted to operators $\hat{\mathcal{J}}_X$ acting on the Hilbert space of the unitary irreducible representation associated with the corresponding orbit \cite{Neri:2025fsh}. These are precisely the operators we have simply been calling $X$ in \eqref{discreteseriesrepresentation}. In conclusion, we arrive at the identification \eqref{eq:chargeoperatorcorrespondence} between the charges of JT gravity and the algebra operators acting in the representation \eqref{discreteseriesrepresentation}.

\section{Relative entropy of coherent states}\label{appendix:relent}

Let us look at two coherent states \eqref{cost} in the same representation. For simplicity, we will choose the coherent states defined by 
\begin{Align}
    \ket{\zeta,\gamma_\zeta} &= \intpE{p}{E} \psi_\zeta(E)\phi_{\gamma,\zeta}(p)\ket{E,p},\\
    \ket{\tilde{\zeta},\tilde{\gamma}_{\tilde{\zeta}}} &= \intpE{p}{E} \psi_{\tilde{\zeta}}(E)\phi_{\tilde{\gamma},\tzeta}(p)\ket{E,p}.
\end{Align}
The relative entanglement entropy between the reduced density operator associated with those two states is given by
\begin{Align}
S\Bigl((\zeta,\gamma_{\zeta})\mid \mid (\tilde{\zeta},\tilde{\gamma}_{\tilde{\zeta}})\Bigr) &= \intpE{p}{E}  \abs{\psi_\zeta(E)}^2 \abs{\phi_{\gamma,\zeta} (p)}^2 \ln\qty(\frac{\abs{\psi_{\zeta}(E)}^2 \abs{\phi_{\gamma,\zeta}(p)}^2}{\abs{\psi_{\tzeta}(E)}^2 \abs{\phi_{\tilde{\gamma},\tzeta}(p)}^2})\\
    &=\int \dd E \,\abs{\psi_{\zeta}(E)}^2\ln\qty(\frac{\abs{\psi_{\zeta}(E)}^2}{\abs{\psi_{\tzeta}(E)}^2}) +  \int \dd p \,\abs{\phi_{\gamma,\zeta}(p)}^2\ln\qty(\frac{\abs{\phi_{\gamma,\zeta}(p)}^2}{\abs{\phi_{\tgamma,\tzeta}(p)}^2})
\end{Align}
Using the explicit wave functions \eqref{modulussquarewavefunctioncoherentsate}, we get
\begin{Align}
    S\Bigl((\zeta,\gamma_{\zeta})\mid \mid (\tilde{\zeta},\tilde{\gamma}_{\tilde{\zeta}})\Bigr) &= (1+2s)\qty(\frac{2(\tilde{r}-r)}{(\tilde{r}+1)(r-1)} + 2 \tanh^{-1}(\tilde{r})-2\tanh^{-1}(r)) \\ &\quad +(r-\tilde{r}) + \frac12(e^{2(\tilde{r}-r)}+4 e^{2\bar{r}}\Im(\gamma-\tilde{\gamma})^2-1).
\end{Align}
One can easily check that the above is always non-negative and vanishes only in the case $(r,\gamma)=(\tilde{r},\tilde{\gamma})$. Let us look at a specific example that has illuminating consequences. We consider the classical state $(r,\gamma)=(\tanh(s),0)$ and a state infinitesimally close to it $(\tilde{r},\tilde{\gamma})=(\tanh(s) + \frac{\delta r}{s},0)$.
The relative entropy is then given by
\begin{equation}
    S\Bigl(\psi^{cl}_s \mid \mid \psi^{cl}_s + \delta r/s\Bigr) =\qty(1 + 2\qty(2s+1)\cosh^4(s))\frac{\delta r^2}{s^2} + \mathcal{O}\qty(\delta r^3). 
\end{equation}
We therefore have the following inequality for different representations
\begin{equation}
    S\Bigl(\psi^{cl}_s \mid \mid \psi^{cl}_s + \delta r/s\Bigr) < S\Bigl(\psi^{cl}_{\tilde{s}} \mid \mid \psi^{cl}_{\tilde{s}} + \delta r/\tilde{s}\Bigr), \quad s < \tilde{s}.
\end{equation}
This is consistent with interpreting the representation parameter as the effective size of the system, since the relative entropy always increases with the size of the system that the reduced density operators describe~\cite{Wehrl:1978zz,Vedral:2002zz}. Furthermore, in the black hole treatment (see Section~\ref{sec:arealaw}), the representation parameter is associated with the area of the subsystem. This is also consistent with the interpretation above.
\par The relative entropy is related to the first law of entanglement and to the Fisher information metric. In our case, the relative entropy depends on two parameters, $r$ and $\Re(\gamma)$. Choosing $\gamma$ to be real for simplicity, one can compute the variation
\begin{equation}
    S\Bigl((\gamma+\delta \gamma,r+\delta r)\mid\mid (\gamma,r)\Bigr) = \qty(1+2\frac{2s+1}{(r^2-1)^2})\delta r^2 + 2 e^{2r}\delta \gamma^2+ \mathcal{O}(\delta r^3,\delta \gamma^3).
\end{equation}
The vanishing of the first order variation is the first law of entanglement entropy which in our case gives
\begin{equation}
    \delta S^{(\zeta,\gamma)} =(2s+1) \frac{3-r^2}{1-r^2} \delta r = \delta \expval{K}_{r},
\end{equation}
where $\expval{K}_r$ is the expectation value of the modular Hamiltonian in the perturbed state. We observe here that the modular Hamiltonian has a linear dependency on the representation parameter $s$.
The second order term in the variations gives the Fisher information metric
\begin{equation}
    G_F(r,\gamma) = \qty(1+2\frac{2s+1}{(r^2-1)^2})\delta r^2 + 2 e^{2r}\delta \gamma^2.
\end{equation}

\bibliography{file1.bib}

\end{document}